\documentclass[12pt]{article}
\pdfoutput=1
\usepackage{url}
\usepackage{jheppub} 
\usepackage{simpler-wick}
\usepackage{braket}

\newcommand{\p}{\partial}

\newcommand{\be}{\begin{equation}}
\newcommand{\ee}{\end{equation}}
\newcommand{\al}{\alpha}

\def\nspc{\hspace{-.25mm}}

\newcommand*{\defeq}{\mathrel{\rlap{%
                     \raisebox{0.3ex}{$\m@th\cdot$}}%
                     \raisebox{-0.3ex}{$\m@th\cdot$}}%
                     =}
\makeatother

\def\be{\begin{eqnarray}}
\def\ee{\end{eqnarray}}

\newcommand{\ra}{\rangle}

\newcommand{\bea}{\begin{eqnarray}}
\newcommand{\eea}{\end{eqnarray}}
\def\ben{\begin{equation}}
\def\een{\end{equation}}

\let\a=\alpha    
   \let\k=\kappa
\let\l=\lambda    \let\p=\phi \let\r=v
\let\s=\sigma \let\t=\tau  \let\c=\chi

\let\w=\omega  \let\D=\Delta

\def\be{\begin{equation}}
\def\ee{\end{equation}}
\def\ba{\begin{array}}
\def\ea{\end{array}}

\def\ZZ{\mbox{\small $Z$}}

\def\ba#1\ea{\begin{align}#1\end{align}}
\def\bs#1\es{\begin{split}#1\end{split}}

\renewcommand{\p}{\partial}
\def\smpc{\hspace{.5pt}}

\numberwithin{equation}{section}
\allowdisplaybreaks  

\title{\boldmath $T\bar{T}$ deformed CFT as a non-critical string}

\author[a,c,d]{Nele Callebaut,}
\author[b]{Jorrit Kruthoff}
\author[a]{and Herman Verlinde}

\affiliation[a]{Joseph Henry Laboratories, Princeton University, Princeton, NJ 08544, USA}
\affiliation[b]{Stanford Institute for Theoretical Physics, Stanford University, Stanford, CA 94305, USA}
\affiliation[c]{Department of Physics and Astronomy, Ghent University, Krijgslaan 281-S9, 9000 Gent, Belgium} 
\affiliation[d]{Department of Mathematics and 
	Haifa Research Center for Theoretical Physics and Astrophysics, 
	University of Haifa, Haifa 31905, Israel} 

\abstract{We present a new exact treatment of  $T \bar{T}$ deformed 2D CFT in terms of the worldsheet theory of a non-critical string. The transverse dimensions of the non-critical string are represented by the undeformed CFT, while the two longitudinal light-cone directions  are described by two scalar fields $X^+$ and $X^-$ with free field OPE's but with a modified stress tensor, arranged so that the total central charge adds up to 26. The relation between our $X^\pm$ field variables and 2D dilaton gravity is indicated. We compute the physical spectrum and the partition function and find a match with known results. We describe how to compute general correlation functions and present an integral expression for the three point function, which can be viewed as an exact formula for the OPE coefficients of the $T \bar{T}$ deformed theory. We comment on the relationship with other proposed definitions of local operators.
}

\def\is{\! &  = &  }
\def\spc{\hspace{1pt}}
\newcommand{\es}[2] {\begin{equation} \label{#1} \begin{split} #2 \end{split} \end{equation}}

\begin{document} 
\maketitle

\pagebreak
\section{Introduction}

\vspace{-2mm}

\addtolength{\abovedisplayskip}{1mm}
\addtolength{\belowdisplayskip}{1mm}
\addtolength{\parskip}{.5mm}
\addtolength{\baselineskip}{.2mm}
\def\fmmu{{}}
\def\mmu{{}}
\def\al{\alpha}
\def\ov{\over}

\def\eepsilon{\mbox{\small ${\cal E}$}}

\def\li{|}
\def\ra{\rangle}

Recently, there has been active interest in the irrelevant deformation of 2D conformal field theory obtained by turning on an irrelevant $T \bar{T}$ coupling
\bea
\label{ttbardeform}
\frac{\partial}{\partial \mu} S_\text{QFT}(\mu) = \int \! d^2x\ \mathcal{O}_{T\spc \bar T},
\eea
with $\mathcal{O}_{T\spc \bar T} = \tfrac{1}{8}(T_{\alpha\beta}T^{\alpha\beta} - (T^\alpha_\alpha)^2)$ and $S_{\rm QFT}(0) = S_{\rm CFT}$. This deformation was  introduced by Zamolodchikov \cite{Zamolodchikov:2004ce} and further studied by Zamolodchikov and Smirnov \cite{Smirnov:2016lqw} and others \cite{Cavaglia:2016oda, MMV, dubov, Giveon:2017nie, Kraus:2018xrn} as a non-trivial continuous family of two-dimensional quantum field theories with strongly coupled UV dynamics. In particular, they showed that due to special integrability properties and a remarkable factorisation formula for expectation values of the  $T\spc \bar T$ operator, many observables become exactly calculable in the deformed theory. 

The partition function and correlation functions of the deformed theory can be formally defined at finite coupling by integrating the flow equation
\bea
\label{ttbflow}
\frac{\partial }{\partial \mu} \bigl\langle {\cal O}_1(x_1)\ldots {\cal O}_n(x_n) \bigr\rangle_\mu \is\Bigl\langle \int\! d^2z\,  T\spc \bar T\,  {\cal O}_1(x_1)\ldots {\cal O}_n(x_n) \Bigr\rangle{}_{\strut \! \mu}.
\eea
This irrelevant interaction is non-renormalizable and expected to destroy short distance locality. Indeed, it is not clear how to use the above flow equation in practice without running into UV divergences. 
Nonetheless,  there are indications that the deformed CFT defined by \eqref{ttbardeform} represents a consistent unitary quantum theory. Besides the fact that the deformation appears to define an integrable theory with an infinite set of conserved charges, concrete supporting evidence for its existence as a well-controlled theory is provided by its relationship with the Nambu-Goto (NG) string in the special case that the undeformed CFT has central charge $c=24$ \cite{Cavaglia:2016oda, MMV}. The goal of this paper is to generalize the equivalence between the $T \bar T$ theory and a Nambu-Goto theory to general values of the central charge. 

Specifically, we will show that the spectrum and correlation functions of the $T\bar{T}$ deformed theory can be mapped to those of the well defined worldsheet theory of a non-critical string. The worldsheet is described by a {\it free} field theory of two scalar fields $X^+, X^-$, representing the light-cone directions of the target space, coupled via the Virasoro constraints to the unperturbed CFT, which in turn can be thought of as representing the transverse target space directions. As we will see, the $X^+, X^-$ fields will continue to behave like free fields for general central charge $c$, except that the energy momentum tensor now contains an extra term proportional to $\kappa = c/24 -1$. Remarkably, the method for computing the correlation functions of the non-critical string was already developed in the early days of string theory, in the context of the study of light-cone gauge formalism for computing string scattering amplitudes. 
Using a more modern BRST formalism, we also give an explicit characterization and construction of the physical states and operators in the theory.
 Since the scalar fields and reparametrization ghosts are described by free field theory, our non-critical string formulation amounts to an exact non-perturbative solution of the  $T\bar{T}$ deformed theory: the deformed amplitudes are obtained in a computable way from the correlation functions of the undeformed CFT. 
 
 As a concrete example, we will compute the matrix element of a local operator $\widetilde{\cal O}_{h}$ in the deformed theory between two states with incoming energy and momentum $(\eepsilon,J)$ and outgoing energy and momentum $(\eepsilon', J')$. The result takes the form of an integral over the cross ratio of a four-point conformal block of the undeformed CFT, smeared against some known function $f(\rho)$
 \bea 
\langle \eepsilon',J' | \widetilde{\cal O}_{h}
| \eepsilon, J \rangle  \; = \;  
\int_0^1\!\! d\rho \, f(\rho) \,\bigl\langle {\cal O}_{\mbox{\tiny ${\nspc \Delta'_L}$}}\!(\infty) \spc {\cal O}_{\mbox{\tiny ${\nspc \Delta_L}$}}\! (0) \spc {\rm P}_{h,p} \spc {\cal O}_{\mbox{\tiny ${\nspc \Delta_R}$}}\!\spc(1) \spc {\cal O}_{\mbox{\tiny ${\nspc \Delta'_R}$}}\!(1-\rho)
\bigr\rangle  
\eea
equal to a correlation function of the $X^+,X^-$ theory. The calculation and the explicit form of the function $f(\rho)$  are described in section 5.

This paper is organized as follows. In section 2, we introduce the worldsheet of the non-critical string theory, and verify its consistency
and its equivalence with the $T \bar T$ deformed CFT at arbitrary central charge.  We present an explicit dictionary between the non-critical worldsheet theory  and the CFT to 2D dilaton gravity, and write an explicit formula for correlation functions of the $X^+,X^-$ theory. In section 3, we compute the spectrum and the thermal partition function. In section 4, we propose a definition of local operators in the  deformed theory in terms of boundary states and in section 5 we outline the computation of the matrix elements, yielding a formal integral expression for the deformed OPE coefficients.  We comment on the relationship with other proposed definitions of local operators of the $T \bar{T}$ deformed theory.

\smallskip

\section{A non-perturbative definition of $T \bar{T}$}
\vspace{-1mm}
In this section we introduce  the non-critical string world-sheet theory and establish its equivalence with the $T \bar T$ deformed CFT at arbitrary central charge. As a warm-up,
we briefly review the mapping between the $T \bar{T}$ theory for $c=24$ and the world-sheet of the critical string. 
\subsection{$T \bar T$ at $c=24$ as a critical string worldsheet}
Consider a general CFT with central charge $c=24$. We define its $T \bar T$ deformed theory by adding two free massless scalar fields $X^+= X^0 + X^1$ and $X^- = X^1- X^0$. 
The total action reads
\bea
\label{xpmact}
S_\text{QFT} \, \is \, S_\text{CFT}\, + \spc  \frac 1 {2\pi \mu} \int \! d^2z \, \bar\p X^+ \spc \p  X^-. 
\eea
The light-cone scalar fields decompose into chiral halves
\bea
X^+\!(z,\bar z)\spc =\, X^+\!\spc(z) + \tilde X^+\!\spc (\bar z), \qquad X^-\!\spc(z,\bar z) \spc = \spc  \tilde{X}^-\!\spc(z) +  X^-\!\spc (\bar z).
\eea
We assume that the spatial target space direction is compactified to a circle $X_1 \equiv X_1 + R$, and restrict to the winding number one sector. Using radial quantization coordinates, we impose~that
\bea
\label{winding} X_1\bigl(e^{2\pi i } z, e^{-2\pi i }\bar z\bigr) = X_1(z,\bar z) + 2\pi R.
\eea 
In the following, we use natural units: the dimensionful coupling $\mu$ of the $T \bar{T}$ theory, related to the string tension of the critical worldsheet theory \eqref{xpmact} as $1/(2\pi \mu)$, is set equal to 1. The strength of the $T \bar{T}$ coupling is parametrized by means of the radius $R$ of the spatial circle. Taking the weak coupling limit of the $T\bar T$ theory corresponds to sending $R\to \infty$.

We interpret the action \eqref{xpmact} (supplemented by a pair of reparametrization ghosts) as the gauge fixed action obtained from a covariant action with a dynamical 2D metric, after going to the conformal gauge. The equations of motion thus include the Virasoro condition that the total stress tensor of the CFT, the  $X$ fields and reparametrization ghosts must vanish for physical states. Or equivalently, using the old covariant formalism,  the equations of motion derived from the action \eqref{xpmact} are supplemented by the Virasoro conditions
\bea
\label{virconstcrit}
-\p X^+ \p X^- \! +\mmu   \spc T_\text{\,CFT} \is 0\, , \qquad \qquad 
-\bar\p X^+ \bar\p X^- \!  + \mmu  \spc \bar T_\text{\,CFT}\, = \,  0\, ,
\eea
which implement gauge invariance under conformal transformations $(z,\bar z) \to (x^+(z), x^-(\bar z))$.

The equivalence between the NG theory and the $T \bar{T}$ deformed CFT is made most evident by comparing the energy spectra of both theories. We postpone this discussion to section 3. Here we briefly recall a physical argument for the equivalence \cite{NG, MMV}.  We could decide to fix the conformal invariance by choosing worldsheet coordinates $(x^+,x^-)$ such that
\bea
\label{lcgauge}
\p X^+ \, \is \, \bar\p X^- = \, 1\, .
\eea
 This temporal gauge condition identifies the worldsheet coordinates with the chiral halves of the target space light-cone fields:  $X^+(x^+,x^-) = x^+ + \tilde{X}^+(x^-)$ and $X^-(x^+,x^-) = x^- + \tilde{X}^-(x^+)$. The other chiral halves are found by integrating the Virasoro constraints 
\bea
\label{vircot}
\p_+  {X}^-   + \, \mmu \spc T^\text{\,CFT}_{++} \is 0\,, \, \qquad \qquad    
\  \ \ \ \ X^- \!  = \, x^-+  \mmu \int^{x^+}\!\!\!\!\!\! dx^+\spc T^\text{\,CFT}_{++}, \nonumber \\[-2.75mm]   
&& \quad\qquad \implies  \\[-2.75mm]\
  \p_- {X}^+    + \, \mmu\spc T^\text{\,CFT}_{--}  \is 0\, .
\,  \qquad\qquad   \ \ \ \ \ X^+ \! =\,x^+ + \mmu\! \int^{x^-}\!\!\!\!\!\! dx^-\spc T^\text{\,CFT}_{--}\, . \nonumber
\eea
Equations \eqref{vircot} provide the on-shell quantum definition of the light-cone coordinates. The fact that they are expressed in terms of the stress tensor indicates that the space-time geometry is dynamical, and that left- and right-moving modes of the CFT are influencing each others trajectory via a geometric shockwave interaction.

 Note that the periodic boundary condition \eqref{winding} of the target space field $X_1 = \frac 1 2 (X^+ + X^-)$ implies that the light-cone coordinates $x^\pm$ satisfy an operator valued periodic identification \bea
\label{pperiod}
(x^+\!,x^-) \, \simeq \, (x^+ \!\!\spc  + 2\pi R - P_-,\, x^- \!\!\spc  + 2\pi R- P_+) \\[-11mm]\nonumber
\eea
where \bea P_\pm =
\oint dx^\pm T_{\pm\pm}^\text{\spc  CFT}\eea
denote the total CFT light-cone momenta.  
This periodicity condition implies that the temporal gauge formulation of the non-critical string theory leads to a non-trivial interaction between the left- and right-moving sectors: every time a right mover passes through a left mover, it gets shifted by an amount proportional to the light-cone momentum of the other particle. Based on this and the above general description, it is not hard to convince oneself that this interacting theory is indeed equivalent to the  $T \bar T$ deformed CFT. A direct way to see this is to  plug \eqref{virconstcrit} back into the free field action \eqref{xpmact}. This directly reproduces the $T \bar T$ interaction term. 

\subsection{Equivalence to 2D dilaton gravity}
\vspace{-1mm}

Before turning to the generalization of the Nambu-Goto formulation to the non-critical case, it will be useful to first describe an alternative proposed definition of the $T \bar T$ deformed theory \cite{Dubovsky:2017cnj}, obtained by coupling the CFT to a dilaton gravity theory with the action
\bea
\label{daction}
S \is \int d^2 z \, \sqrt{g} \left( \Phi R \, +\,  \frac{2}{\mu} \spc \right) .
\eea
Here $R$ denotes the Ricci scalar curvature of the dynamical metric $g$. The dilaton field $\Phi$ plays the role of a Lagrange multiplier imposing the flatness condition $R=0$. The cosmological constant term, determined by $\mu$, introduces the mass scale that sets the strength of the $T \bar{T}$ coupling. We will again choose units so that $\mu=1$. 

In conformal gauge $ds^2 = e^{2\rho} dz d\bar z$, the action \eqref{daction} becomes
  \bea
  \label{caction}
S \is \frac{1}{2} \int\! d^2 z \bigl(\Phi \hat{R}  -8 \Phi \p\bar\p {\rho}  +  2e^{2 {\rho}} \bigr) \ 
\eea
with $\hat{R}$ the Ricci scalar curvature of a fixed background metric $\hat{g}$. We will often choose this metric to be flat. 
The equations of motion derived from \eqref{caction} then take the form
\bea
\p \bar\p \rho \is 0, \qquad
-2 \p \bar \p \Phi +  e^{2\rho}\, = \, 0 \, .
\eea 
We can thus choose to parametrize the space of on-shell field configurations for ${\rho}$ 
with the help of two chiral scalar fields $X^+(z)$  and $X^-(\bar z)$  via 
\bea
\label{flatm}
ds^2 \is  e^{2\rho(z,\bar z)} dz d\bar z = \p X^+ \bar \p X^- dz d\bar z.
\eea
These are the chiral halves of the worldsheet light-cone coordinate fields of the NG formulation. The other chiral halves $\tilde{X}^+(\bar z)$ and $\tilde{X}^-(z)$  parametrize the on-shell classical field configurations of the dilaton field $\Phi$ via
\bea
\label{omegasol}
\Phi(z,\bar z)\, \is\spc \frac{1}{2} X^+(z) X^-(\bar z) \spc +\spc \w^+(z) \spc +\spc  \w^-(\bar z)\nonumber \\[-2mm]\\[-1mm]
\label{Ppdef}
\p \w^+\! \is  -\frac{1}{2}\tilde{X}^- \p X^+, \qquad \quad \bar \p \w^-\!  = -\frac{1}{2}\tilde{X}^+ \bar \p X^-. \nonumber
\eea
It is not difficult to show that the above field redefinitions specify two conjugate non-chiral scalar fields $X^\pm$ with the free field OPE relation \eqref{wick}. 
\footnote{The field redefinitions \eqref{flatm} and \eqref{omegasol} are in fact directly related to the well-known bosonization rules for the $(\beta,\gamma)$ chiral boson system with spin (0,1). 
Identifying $\partial X^+ = \gamma$ and $\tilde{X}^- = \beta$ and similar for the left movers, we can use the familiar bosonization rules 
\bea
\partial {X}^+ = \eta \spc e^{-\varphi} =  e^{-\chi-\varphi} , \qquad \quad \tilde{X}^- =  \partial \xi e^{\varphi} = \partial \chi \spc e^{\chi+\varphi}, \qquad \quad \tilde{X}^-\partial X^+ = \partial \varphi
\eea
where $\varphi$ is the bosonized $\beta\gamma$ current, $(\xi,\eta)$ are anti-commuting chiral fields with spin (0,1), and $\chi$ is the bosonized $\xi\eta$ current. The dictionary \eqref{flatm}-\eqref{omegasol} then follows via the identification $2\rho(z) =  - \chi -\varphi$  and $2\omega^+ = \chi - \varphi$, etc.
}

Applying this dictionary to the stress tensor of the dilaton gravity theory yields the standard free scalar field stress tensor for the $X^\pm$ fields \cite{Verlinde:1993sg},
\bea
T_{\rm grav}\, \is\spc - 4 \p \Phi\p{\rho} + 2\p^2 \Phi  \, = \, - \partial X^+ \partial X^-.
\eea
Note that both stress tensors indeed have the same central charge $c=2$. The combined theory of the dilaton gravity with the CFT must satisfy the Virasoro conditions $T_{\rm grav} + T_{\rm CFT} = 0$.

\newcommand{\ali}[1]{\begin{align} #1 \end{align}}

\def\alphaprime{}

\subsubsection{Dilaton gravity at general central charge}

Next let us consider the coupling of the dilaton gravity theory to a CFT with general central charge $c$. To obtain a well-defined quantum theory,  the central charge of the total stress tensor  $T = T_{\rm grav} + T_{\rm CFT} + T_{\rm ghost}$ 
must add up to zero.  However, when combining the CFT with the above naive dilaton gravity theory and ghosts, we end up with a world-sheet theory with total central charge $2 - 26 + c = c-24$.
So for a CFT with $c \neq 24$, we must account for the mismatch and cancel the conformal anomaly by adding a Liouville action
\bea
\label{laction}
\kappa S_{\rm L}(\rho) \is \frac{\kappa}{2}\int d^2 z  \, \sqrt{g}  \, {R} \, 
\frac{1}{{\square}}\, 
{R} \, = \, -{\kappa} \int \!d^2 z\, \bigl(4 \p {\rho} \bar{\p} {\rho} + \hat R \rho\bigr)
\eea
where $\kappa = \frac c {24} - 1$.  Alternatively, we can think of this action $S_L$ as part of the CFT effective action in the dynamical background $ds^2 = e^{2\rho} dz d\bar{z}$.  Note that the Liouville action \eqref{laction}  (naively) vanishes on-shell. However, its presence will still affect the form of the stress tensor. 

Defining $\Omega = \Phi - \kappa \rho$,  the total action, the equations of motion for $\hat R = 0$  and the stress tensor take the form
  \bea
  \label{omegaaction}
S \is \frac{1}{2} \int\! d^2 z \bigl( \hat{R} (\Omega - \kappa\rho) - 8 \Omega\p\bar\p {\rho} +  2 e^{2 {\rho}} \bigr) \, ,
\\[3mm]
& & \p \bar\p \rho \spc = \spc 0, \quad\
-2 \p \bar \p \Omega+  e^{2\rho}\, = \, 0 \, , \\[3.5mm]
\label{tdil}
& & T\, = \spc - 4 \p \Omega \p{\rho} + 2\p^2 \Omega - 2 {\k} \p^2 \rho\, . 
\eea
The stress tensor of the dilaton gravity theory now has central charge $2- 24\kappa = 26-c$, as required for consistency. Note that the new dilaton field $\Omega$ no longer transforms as an ordinary scalar field, but the combination $\Phi = \Omega + \kappa \rho$ still does.

How does this modification of the dilaton gravity theory translate to the $X^\pm$ scalar field theory? The relation \eqref{flatm} between $\rho$ and the chiral halves of $X^\pm$ remains in place. The classical solution of $\Omega$ remains identical to that of $\Phi$ given in \eqref{omegasol} \footnote{An alternative theory could be written down from the action \eqref{omegaaction} in terms of $\Phi$ and $\rho$,  by using the equation \eqref{omegasol} for $\Phi$ (rather than $\Omega$). In such a theory, see for example \cite{Baba:2009ns}, the fields $\tilde X^+(\bar z)$ and $\tilde X^-(z)$ have a non-trivial OPE with themselves.}  
and the relation between $\omega^\pm$ and the other chiral components of the scalar fields is also unchanged. The only change is that the form of the stress tensor of the $X^\pm$ theory gets modified, as described below.

In the following,  we will choose to work with the $(X^+,X^-)$ formulation, because it makes the (target) space time symmetries (Lorentz invariance and translation invariance) more manifest. As we  will see, it will allow us to define operators that are localized within the target space and use them to study the space-time properties of the $T\bar{T}$ theory more directly.

 \medskip
 
\subsection{$T \bar T$ as a non-critical string worldsheet}

Applying the above dictionary \eqref{flatm}-\eqref{omegasol} to the non-critical dilaton gravity theory described above, we arrive at the following new description of a non-critical worldsheet theory. 

The non-critical worldsheet theory consists of a general CFT coupled to two free scalar fields $X^+,X^-$ satisfying the usual free field Wick contraction rule
\bea
\label{wick}
	\wick{\c {\p X^+}\!\spc(z) \c {\p X^-}\!\spc(w)} \, \is  \frac{-\, 1\ \ }{(z-w)^2} \qquad , \qquad 
	\wick{\c {\p X^\pm}\!\spc(z) \c {\p X^\pm}\!\spc(w)} \, =\, 0\, .  
\eea
The stress energy tensor of the light-cone scalar fields takes the following form
\bea
\label{tnew}
T \is -\, \p X^+ \p X^- \! - \, \kappa \,\p^2 \! \log \p X^+, \nonumber\\[-2.5mm]\\[-2.5mm]\nonumber
 \bar T \is  - \, \bar\p X^+ \bar\p X^-\! -\, \kappa \, \bar\p^2 \! \log \bar\p X^-.
\eea
Using \eqref{wick}, it is a straightforward calculation (see appendix \ref{appA}) to show that this stress tensor satisfies the standard OPE relation
\bea
	T(z) \, T(w)\, \is \,  \frac{c_X/2}{(z-w)^4} + \frac{2 T(w)}{(z-w)^2} + \frac{\p T(w)}{z-w}\;  + \;\ldots    \label{TTOPE}
\eea
with central charge $c_X = 2 - 24 \kappa$.
For our non-critical string worldsheet, we will therefore choose
\bea
c + c_X - 26 = 0, \qquad & \leftrightarrow & \qquad \kappa\, = \frac{c}{24} - 1.
\eea
The stress tensor \eqref{tnew} looks perhaps somewhat unfamiliar and unpractical, but it directly arises by applying the free field paramatrization \eqref{flatm}-\eqref{omegasol}  to the non-critical 2D dilaton gravity theory defined in the previous section. As we will see, because the extra term in the stress tensor does not have an OPE with itself, the modification behaves in a rather well mannered way. It has appeared in previous literature \cite{svv, Chung:1993rf} in the context of studies of the CGHS model \cite{PhysRevD.45.R1005, Russo:1992ax}. Indeed this model can be recast into the form \eqref{daction} through a Weyl transformation.

The free field property \eqref{wick} of the scalar fields and the stress tensor \eqref{tnew} are perfectly compatible, and can both be derived from a well-defined worldsheet action
\bea
\label{newaction}
S_X \is \frac{1}{2\pi} \int \! d^2z \, \Bigl( \bar\p X^+ \spc \p  X^-\! -\, \frac{\kappa}{4} \hat{R} \log (\p X^+ \bar\p X^-)\bigr) .
\eea
The stress tensor, defined via the variation of the action with respect to worldsheet metric, takes the non-standard form \eqref{tnew}. Note that the extra term can be written as a linear dilaton term $\tfrac{1}{2}\kappa \hat{R} \rho$ with $\rho$ defined as the conformal factor of the flat metric $e^{2{\rho}} = \p X^+ \bar\p X^-$ parametrized by the light-cone scalar fields. 
The extra $\kappa$ term in the stress tensor \eqref{tdil} of the dilaton gravity theory maps to the extra $\kappa$ term in \eqref{tnew}.
Indeed, we claim that the $X^{\pm}$ theory is equivalent to the dilaton gravity theory \eqref{omegaaction}, c.f. \cite{dubov18}. The action \eqref{newaction} preserves 2D Poincar\'e invariance in the 2D target space. In particular, we are allowed  to impose the periodicity condition  \eqref{winding}.

The Weyl and conformal anomalies of the combined theory of the CFT, the $X$-fields and ghosts vanish. We could use the Weyl invariance to choose a flat background metric $d\hat{s}^2 =\, dz \spc d\bar z$ with vanishing Ricci scalar   $\hat{R} \, = \, 0$ within any local region. 
The action \eqref{newaction} then  reduces to the standard gaussian free field action \eqref{xpmact}. We can therefore compute local properties, such as the operator products, by applying the usual free field Wick contraction rules \eqref{wick}.  Note however that, while $X^+(z)$ and $X^-(\bar z)$ transform as normal chiral scalars, the other chiral halves $\tilde{X}^+(\bar{z})$ and $\tilde{X}^-(z)$ have a non-standard transformation rule under conformal transformations. The existence of Weyl invariance and the free field Wick contractions are important ingredients to understand correlation functions in the $X^{\pm}$ theory, to which we now turn.

\subsection{Correlation functions in the $X^{\pm}$ theory}\label{Xpmcorrelators}

We will be interested in correlation functions in the presence of  plane wave vertex operators, 
\bea
 e^{ip_i\cdot X(z_i,\bar{z}_i)} \equiv e^{i(p_{iL}\cdot X(z_i)+p_{iR}\cdot \bar X(\bar{z}_i))}.
 \eea
 In the critical theory, these  are computed by standard free field theory.
The idea behind the computation for the non-critical theory is to collect the contributions due to the conformal anomaly, while using a world-sheet metric that is flat everywhere, except at special locations.
 
 On a topologically trivial surface, the correlation functions of plane wave operators factorize into the product of two chiral components, 
 \bea
 \bigl\langle\,
 e^{ip_{1} \cdot  X(z_1,\bar{z}_1)} ... \,  e^{ip_{n} \cdot X(z_n,\bar{z}_n)} \bigr\rangle \, \is \,
 \bigl\langle\,
 e^{ip_{1L} \cdot X(z_1)} ... \,  e^{ip_{nL} \cdot X(z_n)} \bigr\rangle 
 \bigl\langle\,e^{ip_{1R} \cdot \bar{X}(\bar{z}_1)} ... \,  e^{ip_{nR} \cdot \bar{X}(\bar{z}_n)} \bigr\rangle
 \eea
Correspondingly, we introduce the light-cone coordinates $\bigl(x^+(z),x^-(\bar{z})\bigr)$ by means of the expectation value in the presence of the plane wave vertex operators,
\bea
\partial x^+(z)  \equiv \langle \partial X^+(z) \rangle \is - \frac{i}{2} \sum_{i=1}^n \frac{p_{iL}^+}{z- z_i}\,, \quad \quad 
\bar\partial x^-(\bar{z})  \equiv \langle \partial X^-(\bar{z}) \rangle = - \frac{i}{2}  \sum_{i=1}^n \frac{p_{iR}^-}{\bar{z} - \bar{z}_i} ,
\eea
with the $p_{iL}^+$ and $p_{iR}^-$ satisfying energy-momentum conservation. In the temporal gauge, we choose these chiral expectation values as our light-cone coordinates. In the early string theory literature, the coordinate transformation from the complex coordinate $z$ to the light-cone coordinates $(x^+,x^-)$ is usually referred to as the Mandelstam map.

In the local flat Weyl frame, we can use a free field Wick contraction prescription to compute the OPE of the stress tensor and a plane wave operator. This is worked out in appendix \ref{appA}. The result is simple,
\bea
\label{tshift}
T(z)  \, \, e^{ip_{1L} \cdot X(z_1)} \, \simeq \, \left(\spc \frac{\, \frac 1 4 {p_{1L}^2}\! - \kappa\; }{(z-z_1)^2}\, +\,  \frac{\partial_{z_1}}{z-z_1} \spc \right)  \, e^{ip_{1L} \cdot X(z_1)}. 
\eea
This OPE relation shows that even with the modified stress tensor \eqref{tnew}, the plane wave operators continue to be well behaved primary fields  but that (relative to the critical theory with $\kappa =0$) the conformal dimension gets shifted by $-\kappa$, from $p_L^2/4$ to  $p_L^2/4-\kappa$. 

As our local flat Weyl frame we choose the standard flat metric $ds^2 = dx^+dx^-$ in $x^{\pm}$ coordinates. This metric takes the form $ds^2 = \partial x^+ \bar{\partial} x^- dzd\bar{z}$ in $(z, \bar{z})$ coordinates. The latter metric is not well-behaved at the location of the vertex operators, since $\partial x^+$ and $\bar{\partial} x^-$ diverge at $z = z_i$. This indicates that, in the standard flat metric,  each vertex operator creates an infinite cylindrical tube. Correspondingly, there must also be interaction points at which the tubes split or join. These interaction points correspond to the locations where $\partial x^+$ and $\bar{\partial} x^-$ vanish. Their locations $z=z_I$ and $\bar{z}=\bar{z}_I$ are found by solving the equations
\bea\label{sol}
\partial x^+(z_I) \is - \frac{i}{2} \sum_{i=1}^n \frac{p_{iL}^+}{z_I - z_i} = 0,\qquad\quad \bar\partial x^-(\bar{z}_I)\, =\, - \frac{i}{2}  \sum_{i=1}^n \frac{p_{iR}^-}{\bar{z}_I - \bar{z}_i} = 0.
\eea
 In the light-cone diagrams, these are the points along the diagram where the strings interact.  From the equations \eqref{sol} it is clear that there are $n-2$ of such interaction points $(z_I, \bar{z}_I)$. The index $I$ will thus run from $1$ to $n-2$. 

The two types of degenerations of the $x^{\pm}$ coordinates correspond to curvature singularities of the light-cone diagram:  the locations of the vertex operators carry one unit of positive curvature and each interaction point carries one unit of negative curvature. The Weyl rescaling from the metric $ds^2 = dz d\bar{z}$ to $ds^2 = \partial x^+ \bar{\partial} x^- dzd\bar{z}$  thus produces a non-trivial Liouville action with pre-coefficient $c + 2 - 26 = 24 \k $.
Taking this Liouville action into account will produce the correct scaling properties of the correlators of vertex operators of the $X^{\pm}$ fields. Fortunately, the relevant computation is well documented in the old string theory literature, see for example \cite{Mandelstam:1985ww, green1988superstring}.  The result combines into an overal factor $\mathcal{M} = \mathcal{M}_L\mathcal{M}_R$, where
\bea\label{correction}
\mathcal{M}_L \is 
\Bigl(\, \prod_{i=1}^{n} p_{iL}^+ 
\Bigr)^{-\k } \Bigl( \sum_{i=1}^n p^+_{iL} z_i \Bigr)^{2\k}\Bigl(\, \prod_{I=1}^{n-2} 
\partial^2 x^+ 
(z_I)
\Bigr)^{-\k/2}\nonumber\\[-2.5mm]\\[-2.5mm]\nonumber
\mathcal{M}_R \is 
\Bigl(\, \prod_{i=1}^{n} p_{iR}^- 
\Bigr)^{-\k } \Bigl( \sum_{i=1}^n p^-_{iR} \bar z_i \Bigr)^{2\k}\Bigl(\, \prod_{I=1}^{n-2} 
\bar{\partial}^2 x^- 
(\bar{z}_I)
\Bigr)^{-\k/2}
\eea
We refer to the original literature \cite{Mandelstam:1985ww} for a detailed derivation of this result. 
Notice that here we have chosen a normalization so that as $\k = 0$, $\mathcal{M} = 1$.

By Taylor expansion and \eqref{sol}, the relation between the coordinate $x^+$ and the smooth coordinate $z$ in the neighborhood of the interaction point is of the form $x^+ - x^+(z_I) \sim (z - z_I)^2$. This means that the target space is a double cover of the worldsheet close to these interaction points. This is precisely what we expect from interaction points as strings join and split there. Thus we can think of these interaction points as the insertion of a $\mathbb{Z}_{m=2}$ twist operator with conformal dimension (including the ghost contribution) $\D = \k (m-1/m) = \frac{3}{2}\k$. Let us verify that this also follows from $\mathcal{M}_{L,R}$ by counting its target space scaling, i.e. powers of the momenta. The $\partial^2 x^+$ is linear in the momentum. Collecting the other powers in $\mathcal{M}_L$, we find the total scaling to be $\frac{3}{2}\k (n-2)$. This is consistent with the above interpretation of $\mathbb{Z}_2$ twist fields being located at each of the $(n-2)$ interaction points. 

The total correlator of vertex operators in the $\k \neq 0$ theory is now given by the factor $\mathcal{M}$ multiplied by the free field correlator. We will denote this free field result as the $\k = 0$ form of the correlator. The chiral part of the correlation function of vertex operators in the $\k \neq 0$ theory is thus given by
\bea
\bigl\langle{e^{ip_{1L} \cdot X(z_1)} \dots  e^{ip_{nL} \cdot X(z_n)} }\bigr\rangle_\kappa \is  \bigl\langle{e^{ip_{1L} \cdot X(z_1)} \dots e^{ip_{nL} \cdot X(z_n)}}\bigr\rangle_{\kappa = 0} \; \mathcal{M}_L \,\, .
\eea  
Explicitly, for $n = 2,3$ this gives 
\bea
\bigl\langle{e^{ip_{L} \cdot X(z_1)} e^{-ip_{L} \cdot X(z_2)} }\bigr\rangle_\kappa \, = \, {z_{12}^{-p_L^2/2 + 2\k}},\qquad \qquad\qquad\ \  \\[3.75mm]
\bigl\langle{e^{ip_{1L} \cdot X(z_1)} e^{ip_{2L}\cdot X(z_2)} e^{ip_{3L}\cdot X(z_3)} }\bigr\rangle_\kappa \,= \,\left(p_{1L}^+p_{2L}^+ p^+_{3L} \right)^{-\k/2} 
\prod_{i<j} \, z_{ij}^{p_{iL}\cdot p_{jL}/2 + \k},
\eea
with $p^+_{3L} = -p_{1L}^+ - p_{2L}^+$.  The scaling exponents in these expressions are consistent with the $\kappa$ shift \eqref{tshift} of the scaling dimensions of the vertex operators.  

We will use the above general result to define correlation functions of the $T \bar{T}$ deformed theory.  Before turning to this computation, we will now first show that the theory $S_X$ in \eqref{newaction} coupled to an arbitrary seed CFT has the same spectrum and partition function as the $T\bar{T}$ deformed theory. 

\section{Spectrum and Partition Sum}\label{sec:spectrum}
\vspace{-1mm}

The equations of motion derived from \eqref{xpmact} are supplemented by the Virasoro conditions
\bea
\label{virconst}
-\p X^+ \p X^- \! - \kappa \,\p^2 \log \p X^+ \, +\, \mmu   \spc T_\text{\,CFT} \is 0\, .
\eea
The self-consistency of this constraint theory is well established for  the critical theory with $\kappa =0$ via the no ghost theorem for the critical NG string.  The key result in this proof is the construction of the so-called DDF operators, summarized and generalized to arbitrary central charge $c$ in section 4.2.
 For our context, it demonstrates that the $T \bar T$ deformed CFT is a well defined quantum theory, in which all Hilbert states have positive norm. We will use elements of the proof of the no ghost theorem later on.  

While useful for establishing the equivalence with the $T \bar{T}$ theory,  the temporal gauge \eqref{lcgauge} does not automatically give rise to a well controlled description of the quantum theory. For this purpose, it is more convenient to use a covariant BRST formalism.  The action \eqref{xpmact} has a nilpotent BRST symmetry and corresponding nilpotent charge $Q_{\rm brst} = Q + \bar{Q}$ with
\es{brstcharge}{Q & \, = \, \oint \! dz\,  \Bigl(\spc c \spc \bigl( T_\text{CFT} +  \fmmu\spc T_X  \, + \, \frac 1 2\spc T_{\rm gh} \bigr)\Bigr) \, .}
Physical states are specified by restricting the space of all states to the BRST cohomology
\es{qinv}{Q_{\rm brst} | {\rm phys} \rangle = 0, \qquad & \qquad |{\rm phys} \ra\;  \simeq \; |{\rm phys} \rangle \, + \, Q_{\rm brst} | {\rm *} \rangle,\\[2mm]
\bigl[Q_{\rm brst}, {\cal O}_{\rm phys} \bigr]  = 0, \qquad & \qquad\; {\cal O}_{\rm phys}\;  \simeq\; {\cal O}_{\rm phys} \, + \, 
[Q_{\rm brst}, {\rm *} ],}
defined as the BRST invariant states modulo BRST exact states.
The four equations \eqref{xpmact}, \eqref{winding}, \eqref{tnew}, \eqref{brstcharge} and  \eqref{qinv} provide a complete non-perturbative definition of the $T \bar T$ deformed CFT on a cylinder.

The space of physical states takes the same form as for the critical string, with only some minor modifications. In particular, the non-trivial BRST cohomology is found in the Hilbert space sector with ghost number $-1$. Using radial quantization, the $-1$ ghost vacuum is defined by acting with the product $c(0)\bar{c}(0)$ of the left- and right c-ghosts on the $SL(2,\mathbb{R})$ invariant vacuum. 
Let ${\cal O}_{\Delta}(z,\bar z)$ with $\Delta=(\Delta_L,\Delta_R)$ denote a CFT primary operator with left and right conformal dimension $\Delta_L$ and $\Delta_R$. We can associate to ${\cal O}_{\Delta}$ a physical state with given energy momentum $p$ via (here $|0\ra$ denotes the $SL(2,\mathbb{R}$) invariant vacuum)
\bea
\label{phystate}
\bigl|\spc \Delta,   p\spc \bigr\ra\, \is \,c(0)\bar{c}(0)\spc {\cal O}_\Delta(0)\spc e^{ip\spc \cdot X(0)}\, \bigl|\spc 0\spc \bigr\rangle,
\eea
provided that $p$ satisfies the usual on-shell condition such that $(L_0-1) | \Delta,   p\ra  = (\bar{L}_0-1)| \Delta,   p\ra =0$. Here $L_0 = L^X_0 + L^{\rm CFT}_0$, etc.

\subsection{Spectrum}

\vspace{-1mm}

To obtain the physical spectrum, we need to pay attention to the precise form of the kinetic term of the light-cone fields. In \eqref{xpmact}, the $X$ action looks like that of a 2D string worldsheet 
\bea
\label{ssigma}
S_X \is \frac{1}{2\pi} \int \! d^2z\, (G_{ab} + B_{ab})\spc \p X^a \bar\p X^b,
\eea 
in a constant target space metric $G_{ab} = \eta_{ab}$ and  anti-symmetric tensor field $B^{ab} = B\spc \epsilon^{ab}$ with $B=1$.
In a general background, the chiral momentum zero modes $p^a_{L,R}$ of the scalar fields, defined as the leading term in the mode expansion
\bea
\p X^a \is  - \frac{i}{2} \frac{p^a_{L}}{z}  -  \frac{i}{\sqrt{2}} \sum_{n\neq 0} \a^{a}_n z^{-n-1}, \qquad \bar\p X^a\spc = \spc - \frac{i}{2} \frac{p^a_{R}} {\bar z}  -   \frac{i}{\sqrt{2}} \sum_{n\neq 0}  \bar{\a}^{a}_n \bar{z}^{-n-1},
\eea
decompose into a sum of target space momenta $p^a$ and winding zero modes $w_a$, via 
\bea
p^a_{L,R} = \bigl(p^a \pm   \spc(G^{ab} \pm B^{ab})w_b\bigr).
\eea
In our case, the time like direction $X^0$ is non-compact, whereas the spatial direction $X^1$  is compactified on a circle, as indicated in equation \eqref{winding}. The time-like momentum (= energy) is therefore continuous, $p^0 \in \mathbb{R}$,
and the timelike winding number vanishes, $w_0=0$. The spatial momenta and winding zero modes are both non-zero and quantized via $p^1\!\spc = {n}/{R}$  and $w_1 = m R \spc , \ \ n,m\in \mathbb{Z}.$

The $T \bar T$ deformed CFT corresponds to the sector with winding number one $w_1 = R$. In this sector, we can parametrize the chiral momentum zero modes via
\bea
\label{param}
p^0_{L} = p^0_{R} = {\cal E} +  B R, \qquad p^1_{L} = \frac{J}{R} + R  , \ \ \ p^1_{R} = \frac{J}{R} - R,
\eea
with ${\cal E}$ the target space energy and $J$ the integer (angular) momentum along the spatial circle. Here we kept the $B$-field parameter $B$. As we will see, it is usually set equal to its critical value $B=1$. 

The on-shell condition $L^{\rm \!\spc tot}_0\! -\bar{L}^{\rm \! \spc tot}_0\!=0$ equates $J$ with the difference of the right- and left-moving scale dimension of the CFT primary state
\bea
\Delta_R - \Delta_L \, = \, \frac{1}{4}p_{1L}^2 - \frac{1}{4}p_{1R}^2 \, \is \, J. 
\eea
The other on-shell condition reads 
\bea
L^{\rm\!\spc tot}_0 + \bar{L}^{\!\spc \rm tot}_0 - 2 \, \is -  \frac{1}{2}p_0^2 + \frac{1}{4}p_{1L}^2 + \frac{1}{4}p_{1R}^2  +  \Delta_L + \Delta_R  - 2\kappa - 2 = 0.
\eea
Inserting the parametrization \eqref{param}, this condition results in the following relation between the energy ${\cal E}$ of the deformed theory and the energy $\spc E \equiv \frac 1 {R}( \Delta_L + \Delta_R  - \frac{c}{12})$ of the unperturbed CFT
\bea
 - 2 \left(\frac {\cal E} 2+ \frac {B R} 2 \right)^2 + \frac{{J}^2}{2R^2} + \frac {R^2} 2 +   R\spc E\,  = \, 0.
\eea
The solution to this equation coincides with the spectrum of the $T\bar T$ deformed CFT
\bea
\label{newe}
{\cal E} \is  R \left(-B+ \sqrt{1+ \frac{2\smpc E}{R}+ \frac{J^2}{R^4}}\, \right) \, .
\eea
This result generalizes the known match between the spectrum Nambu-Goto theory and $T \bar T$ deformed CFT to the case of general central charge $c$, provided we set $B=1$.\footnote{From the non-critical string perspective, it is in fact natural to include the $B$-field as an additional tunable coupling of the deformed CFT. The presence of the B-field was also emphasized in \cite{Hashimoto:2019wct}. }  
\medskip

\subsection{Partition sum}

\vspace{-1mm}

Via the match of the spectrum, we are in principle assured that the thermal partition functions of the non-critical string and the $T\bar T$ theory both match. It is still instructive to see how this works in practice from a functional integral perspective, c.f. \cite{dubov2, Hashimoto:2019wct}. We will encounter several subtleties.

Consider the generalized thermal partition 
\bea
\label{zone}
Z(\alpha, \beta) = \sum_{n} e^{ i\alpha \spc  J_n -\beta \spc {\cal E}_n},
\eea
with inverse temperature $\beta$ and  chemical potential $\alpha$ for the spatial momentum $J_n$.
We can represent $Z(\alpha,\beta)$ as the partition function of the $T \bar T$ deformed CFT on a Euclidean two torus. We will parametrize the 2-torus by means of a complex coordinate
$x = x_1 + \sigma x_0$, where we assume that the real coordinates $x_0$ and $x_1$ are both periodic with $2\pi R$. Here $\sigma$ is a complex number, with real and imaginary part related to $\alpha$ and $\beta$ via
$\alpha = 2\pi R \sigma_1$ and $\beta = 2\pi R \sigma_2$.  

It will be helpful to introduce the notation
\bea
\label{lambdas}
\Lambda \, =\,  2\pi R\spc \beta \, = \, 4\pi^2 R^2 \sigma_2, \qquad \qquad \lambda = \frac{1}{4\pi^2 R^2} .
\eea 
Here $\Lambda$ defines the volume of the 2-torus, and $\lambda$ defines the coupling constant of the $T\bar T$ deformation. The two are related via $\Lambda = \sigma_2/\lambda$. Modular transformations of the torus act on $\sigma, \Lambda$ and $\lambda$ via
\bea
\label{modtrafo}
(\, \sigma\, ,\,  \Lambda\, ,\,  \lambda\, ) & \to & \Bigl( \, \frac{a \sigma + b}{c \sigma + d}\, ,\, \Lambda \, , \, \frac{\lambda}{|c\sigma + d|^2}\, \Bigr).
\eea
It has been shown in \cite{Cardy:2018sdv,Aharony:2018bad, dubov18} that the partition function $Z(\alpha, \beta)$, considered as a function of the complex modulus $\sigma$ and coupling constant $\lambda$ 
\bea
Z(\sigma,\bar\sigma, \lambda)\; \is 
\;  \sum_{n}\; e^{2\pi iR\spc (\sigma_1 J_n + i \sigma_2\spc  {\cal E}_n)},
\eea
satisfies the following differential equation
\bea
\label{diffeq}
\frac{d}{d \lambda} \, Z(\sigma,\bar{\sigma},\lambda) \is \left(\sigma_2\spc \p_\sigma \p_{\bar \sigma}  + \frac 1 2 \Bigl( i (\p_\sigma - \p_{\bar \sigma}) - \frac 1 {\sigma_2} \Bigr)\lambda \p_\lambda\right)Z(\sigma,\bar{\sigma},\lambda).
\eea
This exact equation follows quite directly from the flow equation \eqref{ttbflow} defining the $T \bar T$ deformation. In \cite{Aharony:2018bad} it was furthermore shown that, remarkably, this differential equation \eqref{diffeq} follows uniquely from the modular invariance requirement that
\bea
\label{modinv}
Z(\sigma,\bar\sigma, \lambda) \is Z\left( \frac{a \sigma + b}{c \sigma + d},\frac{a \bar\sigma + b}{c \bar\sigma + d}, \frac{\lambda}{|c \sigma + d|^2}\right)\, .
\eea
We will now reproduce this result from the non-critical string formulation. A similar derivation was given previously in \cite{dubov18}.

We need to consider the one-loop partition function of a non-critical string with target space given by a 2-torus. We parametrize the target space torus by means of a complex scalar field 
$X = X_1 + \sigma X_0$, where we assume that the two real scalar fields $X_1$ and $X_0$ are both periodic  with period $1$. (In this subsection, we choose to absorb the period $2\pi R$ into the target space metric.) As before, we identify $\Lambda$ with the volume of the target space torus,
except that (as is standard in the treatment of string worldsheet sigma models) we now combine it with the anti-symmetric two-form field $B_{\mu\nu} = B\epsilon_{\mu\nu}$ into a `complex' volume modulus 
$\Lambda = \frac 1 2  G(1-i B)$ with $G$ the ordinary geometric volume of the $T^2$. 
In this notation, the general sigma model action \eqref{ssigma} takes the form 
\bea
S_X \is \int\! \frac{d^2z}{\sigma_2} \, 
\bigl( \Lambda \spc \p \bar X \bar \p X + \bar \Lambda \spc  \p  X \bar \p \bar X \bigr).
\eea
Comparing this equation with the non-critical string worldsheet action, we see that we need to (formally) set $\bar{\Lambda} =0$, by tuning the anti-symmetric tensor to $B= i$.\footnote{This further requires that we complexify the target space, so that $\Lambda$ and $\bar\Lambda$ become independently tunable.}

The computation of the one loop non-critical string partition sum is now straightforward. The string worldsheet itself is also a torus, parametrized by complex coordinates $z = \xi_1+ \tau \xi_0$ with the real coordinates $\xi_1$ and $\xi_2$ both periodic with period 1. Note that the worldsheet 2-torus admits a flat metric  $d\hat{s}{}^2\! = dzd\bar z$ with $\hat{R}=0$, so we can ignore the extra $\kappa$ term in the world sheet action~\eqref{newaction}. The modular parameter $\tau$ of the worldsheet torus is part of the dynamical worldsheet metric, and needs to be integrated over. The integrand is given by the product of the partition functions of the CFT, the free scalar field $X$, and the ghosts,
all evaluated at a fixed value of $\tau$.
The ghost contribution simply gives the Faddeev-Popov functional determinant $\det \Delta_{\rm FP} = |\det \p|^2 = |\eta(\tau)|^4$. Here $\eta(\tau) = \prod_n (1-q^n)$ with $q= e^{2\pi i \tau}$ is the familiar Dedekind $\eta$-function. This ghost partition function exactly cancels the fluctuation determinant contribution of the complex scalar field $X$. 

The complex scalar fields $X(z,\bar z)$ represent a mapping from the worldsheet torus into the target space torus. We will assume that this mapping has wrapping number one. Conversely, each real scalar field $X_0$ and $X_1$ represents (the inverse of) a mapping of a target space circle (the A-cycle or B-cycle of the target space torus) into the worldsheet torus. The latter mapping is labeled by two winding numbers, which we denote by $w_1 = (m_1, n_1)$ and $w_0 = (m_0,n_0)$, respectively. The remaining partition function thus takes the form of a sum over winding sectors
\bea
\label{zsum}
Z(\sigma,\bar\sigma,\lambda) \is 
 \frac{\s_2}{\pi \l} \int_{\bf F}\! \frac{d^2 \tau}{\tau_2^2}\,\sum_{w} \; e^{-S_{\rm cl}(\Lambda,\spc \sigma,\spc \tau,\spc w)}  \, \,  Z_{\rm CFT}(\tau,\bar \tau),
\eea
where $S_{\rm cl}(\lambda,\spc \sigma,\spc w)$ denotes the classical action of the solution to the classical equations of motion $\p \bar \p X_a^{\rm cl} = 0$ of the scalar field labeled by the winding number $w$.
This classical solution takes the following simple form 
\bea
X_a^{\rm cl}(z,\bar z) \is \frac{1}{\tau - \bar \tau} \bigl((n_a - m_a \bar \tau)\, z  \, + \, (n_a - m_a \tau)\, \bar z\bigr),
\eea
which satisfy the boundary conditions 
\bea
\ X_a^{\rm cl}(z+1,\bar{z} + 1)\  \sim \ X_a^{\rm cl}(z,\bar{z}) + m_a, \nonumber\\[-2.5mm]\\[-2.5mm]\nonumber
X_a^{\rm cl}(z+\t,\bar{z} + \bar{\t})\ \sim\ X_a^{\rm cl}(z,\bar{z}) + n_a.\ 
\eea
Plugging this solution back into the scalar field action, it is now straightforward to show that the sum over winding sectors reduces to
\bea
\label{solsum}
  \sum_{w}\;  e^{-S_{\rm cl}(\Lambda,\spc \sigma,\spc \tau, \spc w)}  \; =\!\!
  \sum_{n_0,m_0,n_1,m_1 \in \mathbb{Z}}\hspace{-7mm}{}^{\mbox{$'$}} \hspace{4.5mm}\exp\left( - \frac{\Lambda } {\tau_2\spc \sigma_2} | n_1 + n_0 \sigma - m_1 \tau - m_0 \sigma \tau| ^2 
  \right).
\eea
Here the $'$ indicates that the sum is restricted to the elementary winding sectors for which
\bea
{\rm gcd}(n_0,m_0) = {\rm gcd}(n_1,m_1) =1, \qquad \quad n_0 m_1 - m_0 n_1 = 1.
\eea
The first restriction implies that each real scalar field $X_a$ has winding number 1, and the second restriction implies that the mapping from the worldsheet torus into the target space torus has wrapping number one. Both restrictions are modular invariant, and the full sum \eqref{solsum} defines a modular expression. We can therefore restrict the integral over $\tau$ in \eqref{zsum} to the usual fundamental domain ${\bf F} = \{| \tau_1 | < \frac 1 2 $, $|\tau| >1\}$. 

Alternatively, we could choose to integrate $\tau$ over the full Poincar\'e upper half-plane ${\bf H}$, and collapse the sum over winding sectors to a single term, say, with $n_0=1$ and $m_1 = 1$. This yields the following final result for the $T \bar T$ partition sum \cite{dubov18, Hashimoto:2019wct}
\bea\label{finalTT}
Z(\sigma,\bar\sigma, \lambda) \is \frac{\s_2}{\pi \l} \int_{\spc \bf H}\! \frac{d^2 \tau}{\tau_2^2}  \, \exp\Bigl(-{\frac{\Lambda} {\tau_2\spc \sigma_2}\, | \sigma - \tau|^2}\Bigr) \, Z_{\rm CFT}(\tau, \bar \tau)\nonumber
\\[-1mm]\label{zfinal} \\[-1mm] \nonumber
\is \frac{\s_2}{\pi \l} \int_{\spc \bf H}\! \frac{d^2 \tau}{\tau_2^2}  \, \exp\Bigl(-{\frac{1} {\lambda \spc \tau_2}\, | \sigma - \tau|^2}\Bigr) \; Z_{\rm CFT}(\tau, \bar \tau).
\eea
Note that this expression manifestly satisfies the modular invariance property \eqref{modinv}: the measure, the quantity in the exponent, and the CFT partition function are all invariant under the simultaneous modular transformations in $\sigma$, $\t$ and $\l$. 

It is straighforward to verify that the integral expression \eqref{finalTT} satisfies the 
differential equation \eqref{diffeq}. Moreover, 
by writing $Z_{\rm CFT}(\tau,\bar \tau) = \sum_n e^{ 2\pi i R( \tau_1 J_n+i \tau_2 E_n) }$, the modular integrals in \eqref{finalTT} can be preformed exactly and it reduces to the earlier definition \eqref{zone} of the $T \bar T$ partition sum with $\mathcal{E}_n(E_n,J_n)$ given in \eqref{newe}. 
It is easy to check that in the $\l \to 0$ limit,
$Z(\s,\bar{\s},\l)$ reduces to the original CFT partition function $
Z_{\rm CFT}(\s, \bar \s).$ Notice that a similar expression was obtained in \cite{Hashimoto:2019wct}, but for the single trace version of the $T\bar{T}$ deformation.

\def\cT{{\mbox{\footnotesize${\cal T}\spc$}}}
\def\cV{{\mbox{\small ${\cal V}\smpc$}}}

\section{Physical operators}
\vspace{-1mm}

We will now turn our attention to the construction of physical operators and correlation functions of the deformed theory.  The existence of a unique notion of deformed local operators is not self-evident, however, since the $T \bar T$ theory is not a standard local QFT with a UV fixed point. In particular,
since the deformation breaks conformal invariance, there is no longer a direct correspondence between states and local operators. 
The best one can do is to identify a natural class of physical (i.e. BRST invariant) operators of the deformed theory,  that in the $\mu \to 0$ limit reduce to a corresponding set of local operators in the undeformed CFT. We first discuss the physical operators that describe the on-shell modes of local operators and the stress tensor, and then we present a proposed definition of the off-shell modes of primary operators of the deformed theory.

\subsection{On-shell operators}
\vspace{-1mm}

On-shell physical operators come in two different types. The first category are local operators of the  form 
\bea
\label{physo}
{\cal O}^{(0)}_{\Delta}(p,z,\bar z)\,  \is\, c(z) \bar c(\bar z) \, 
\mathcal{O}_{\Delta}(z,\bar z)\;  e^{ip\spc \cdot X(z,\bar z)}  
\eea
where $\mathcal{O}_{\Delta}(z,\bar z)$ denotes a primary operator of the undeformed CFT.
For simplicity we assume that $\Delta_L = \Delta_R$ and $p_L=p_R$. So in particular, equation \eqref{physo} describes the on-shell momentum mode of a local scalar operator that acts within a given winding sector of the $X^\pm$ CFT. The operator \eqref{physo} is BRST invariant provided the momentum satisfies the on-shell condition
\bea
\label{onshell}
\frac{-p_0^2 + p_1^2}{4}   + \Delta = \frac{c}{24}.
\eea
A second type of physical operators are given by the integral of the corresponding $(1,1)$ form over the 2D worldsheet. Typically, the $(1,1)$-form operator is obtained by simply stripping off the $c$-ghosts from the local form of the physical operators.\footnote{More formally and generally, given any BRST invariant local operator, one obtains the corresponding $(1,1)$ form version by integrating the descend equations
$[ \smpc Q, \smpc V^{(1,0)} \smpc  ]=\p\spc V^{(0)}$, 
 $[\smpc \bar Q, \smpc {V}^{(1,1)} \smpc  ]=\bar \p\spc V^{(1,0)}$, etc.} We will denote these operators by $\mathcal{V}_{p,\D}(z,\bar{z})$:
 \be
 \mathcal{V}_{p,\Delta}(z,\bar{z}) = \mathcal{O}_{\Delta}(z,\bar z)\;  e^{ip\spc \cdot X(z,\bar z)}.  
 \ee
For the above operator \eqref{physo}, the resulting integrated physical operator takes the standard form of a momentum eigenmode of the CFT operator ${\cal O}_{\Delta}$ 
\bea
\label{vnaive}
{\cal O}(p\spc )\; \is \, \int\! d^2 z \; \mathcal{O}_{\Delta}(z,\bar z)\;  e^{ip\spc \cdot X(z,\bar z)}\;
\eea
along the $X^\pm$ plane. The above construction is adequate for on-shell physical operators, that satisfy the condition \eqref{onshell}. This class is not sufficient to describe an actual local operator. In subsection 4.3, we will present a proposal for an off-shell definition of local operators.

\subsection{Stress tensor}
\vspace{-2mm}

It is instructive to consider the on-shell modes of the stress tensor. They are expressed in terms of contour integrals as follows
\bea
\label{ellen}
\  {\cal L}_n \is  \oint \! dz \, \left( \partial X^- \! +   \, 
i p_+ \hat{\kappa}  \p \log \p X^+\right) e^{i p_+ X^+(z,\bar{z})}, \qquad p_+ = \frac n R,
\eea
with $\hat{\k} = \k + 1/2$. A similar expression holds for the right-moving modes $\overline{\cal L}_n$. 
Equation \eqref{ellen} defines a conserved quantity, provided we work in the purely left-moving sector of the worldsheet CFT: we have to assume that  $P_-$ vanishes, so that left-moving plane waves are periodic with period $2\pi R$. This chiral projection effectively turns off the $T \bar T$ interaction. This explains why conformal symmetry is preserved within this subsector. 

The operators \eqref{ellen} are a direct generalization of the well-known DDF operators of critical string theory, first discovered by  DelGuidice, Divecchia and Fubini. A key property of the DDF operators is that they generate the Virasoro algebra
\bea
\label{viralg}
[{\cal L}_n, {\cal L}_m] \is (n-m) {\cal L}_{n+m} + \frac{c}{12}(n^3-n)\delta_{n+m,0},
\eea 
with central charge $c$ equal to that of the transverse CFT. In view of the earlier discussion of the light-cone gauge formalism in section 2.3,
this is no surprise: via the Virasoro constraints  \eqref{virconst} of the total world-sheet theory, the operators \eqref{ellen} are identified with the  Virasoro generators of the transverse CFT. 
This identification, and the result that \eqref{ellen} satisfies \eqref{viralg}, plays a key role in the original proof of the no ghost theorem of critical open string theory. This classic proof is not available in our case, however, since the generators ${\cal L}_n$ only act within the chiral subsector of the theory. But we can still rely on the modern proof of the no ghost theorem based on BRST invariance \cite{SPIEGELGLAS1987205} to assure ourselves that the physical Hilbert space contains only positive norm states.

\subsection{Off-shell operators}
\vspace{-1mm}

In this section, we will aim for the more ambitious goal of defining off-shell modes of local operators. In particular, we would like to be able to consider correlation functions of CFT operators placed at arbitrary locations on the light-cone cylinder. We thus need to associate physical operators with arbitrary light-cone momenta to each CFT operator $ \mathcal{O}_{h,\bar{h}}$.  

There are different ways to look for such off-shell operators. Schematically, we should look for physical operators of the form
\bea
 c \bar c \, \mathcal{O}_{\Delta}\;   (\p X^+\!\spc)^{1-j} \, (\bar \p X^-\!\spc )^{1-\bar j} \; e^{ip\spc \cdot X}\; +\; \ldots\;
\eea
where $j$ and $\bar j$ are determined via the on-shell condition
\bea
\label{onshell2}
j  =  \Delta_L + \frac{1}{4}p_L^2- \kappa , \qquad & &  \qquad \bar j = \Delta_R + \frac{1}{4}p_R^2 - \kappa \, .
\eea
The $\ldots$ indicate extra terms needed to account for the effect of normal ordering subtractions, that correct for anomalous OPE contributions between the $\p X^+$ and $\bar \p X^-$ factors and the plane wave factor.
In string theory, one typically sidesteps these normal ordering issues by restricting to on-shell vertex operators with $j=\bar j = 1$. Equation \eqref{onshell2} then reduces to the usual mass-shell restriction on the momenta. The normal ordering subtleties reflect the fact that the string worldsheet  is a non-local gravitational system, and the light-cone coordinates $X^+$ and $X^-$ are non-commuting operators that obey a Heisenberg uncertainty relation.  

Here we will instead follow a more determined approach by drawing on lessons from string theory. A local space-time operator in string theory  pins the string worldsheet to a given target space location. The best example of such a local operator is a D-instanton. Inspired by this analogy, and generalizing to include unoriented string worldsheets, we can introduce two types of momentum eigenmodes of local space-time operators as the operators that create boundary states solving the following infinite set of conditions  \footnote{Here $L_n = L_n^{\rm cft} + L_n^{{}^{{}_X}} + L_n^{\rm gh}$ denotes the total Virasoro generator and $\alpha^\pm$ the oscillators of the light-cone scalar fields $X^\pm$.}
\bea
\widehat{\cal O}_{h}(x)|0\ra \is |\!\spc|  h, x\ra\!\ra,\    \qquad\quad\, \,
(L_n\! \spc - \bar{L}_{-n}) |\!\spc|  h,x\ra\!\ra \, =\, (\alpha^{\pm}_n \! - \bar\alpha^{\pm}_{-n})|\!\spc|  h,x\ra\!\ra  \, = \spc  0\ \ \\[3.5mm]
\widetilde{\cal O}_{h}(x)|0\ra \is |\!\spc|  h, x\ra\!\ra_{\otimes},
\qquad\quad
(L_n \! \spc- \widetilde{L}_{-n}) |\!\spc|  h, x\ra\!\ra_{\otimes} =\spc (\alpha^{\pm}_n\! + \widetilde{\alpha}^{\pm}_{-n}) |\!\spc|  h, x\ra\!\ra_{\otimes} =\spc0
\eea
with $\widetilde{L}_{n} = (-)^n\bar{L}_n$ and $\widetilde{\alpha}^{\pm}_{n} = (-)^n\bar{\alpha}^{\pm}_n$. The states $|\!\spc|  h, x\ra\!\ra$ and $|\!\spc|  h, x\ra\!\ra_{\otimes}$ are generalized Ishibashi boundary states of the total worldsheet CFT, built by acting with all possible symmetry generators of the total CFT (defined as the chiral algebra generated by the Virasoro generators and the oscillator modes of $X^+(z)$ and $X^-(\bar{z})$)  on the primary state $|h\rangle_{\rm cft} |x\rangle_{{\!}_{{}^X}} |0\rangle_{\rm gh}$, with $|x\rangle_{{\!}_{{}^X}}$ the ground state with given target space position
\bea
\hat{x}^\pm |x\rangle_{{\!}_{{}^X}}=  x^\pm |x\rangle_{{\!}_{{}^X}}, 
\eea
with $ \hat{x}^\pm$ the constant zero mode of $X^\pm$. 
The operators $\widehat{\cal O}_{h}(x)$ and $\widetilde{\cal O}_{h}(x)$ create a finite size hole or cross cap, respectively,
and thereby add three worldsheet moduli: a position $(z,\bar{z})$  and a scale parameter $y$. The mass-shell operator $L_0+\bar{L}_0 - 2$ does not annihilate the boundary state, but acts by changing the scale parameter. Physical correlation functions are defined by the integral over all moduli, including the location and size of the hole or cross-cap. 
We can introduce momentum eigenstates $\ket{p}_{{\!}_{{}^X}}$ in the $X^{\pm}$ theory by Fourier transforming the position eigenstates: 
\be
\ket{p}_{{\!}_{{}^X}} = \int d^2 x\,e^{ip \cdot x} \ket{x}_{{\!}_{{}^X}} \, . 
\ee
This allows us to give the cross-cap state momentum. 

\medskip

\subsection{Flow equation for local operators}
\vspace{-.5mm}

In recent work \cite{Cardy:2019qao}, Cardy studied the $\mu$ dependence of correlation functions of local operators in the $T \bar{T}$ deformed theory and showed that this dependence can be captured by a flow equation, expressed in terms of line integrals of energy momentum tensors. In this subsection, we will present a schematic derivation of this flow equation starting from our definition of local operators. As in \cite{Cardy:2019qao}, we will consider correlation functions on the infinite plane (or equivalently, we will omit terms in the flow equation that describe the $\mu$ dependence of the initial and final state).

Let us consider an arbitrary $n$-point correlation function of operators $O(x_i)$ in the $T\bar{T}$ deformed theory on the plane. The local operators create circular boundaries of cross caps. With some abuse of terminology, we will refer to these holes or cross caps as `punctures', even though they in fact have a finite size. We would like to study how such correlators change upon changing $\mu$, which we do by taking a $\mu$ derivative of the correlator in the deformed theory. The $\mu$ dependent piece in the action \eqref{xpmact} of the deformed theory is $
S_{\mu} = \frac{1}{2\pi \mu} \int \! d^2 z\, \bar{\partial} X^+ \partial X^-$. 
Differentiating with respect to $\mu$ brings down an insertion of the free action of the $X^\pm$ fields 
\bea
\label{flowCorr}
\frac{\partial \;} {\partial\mu}  \bigl\langle\, {\prod_i O(x_i)}\, \bigr\rangle\is -\frac{1}{2\pi} \bigl\langle{\, \int \! d^2 z \, \bar{\partial} X^+ \partial X^- \, \prod_i O(x_i) \, }\bigr\rangle,
\eea
where on the right-hand side we again took natural units $\mu=1$.

\begin{figure}
\centering
\includegraphics[scale=0.24]{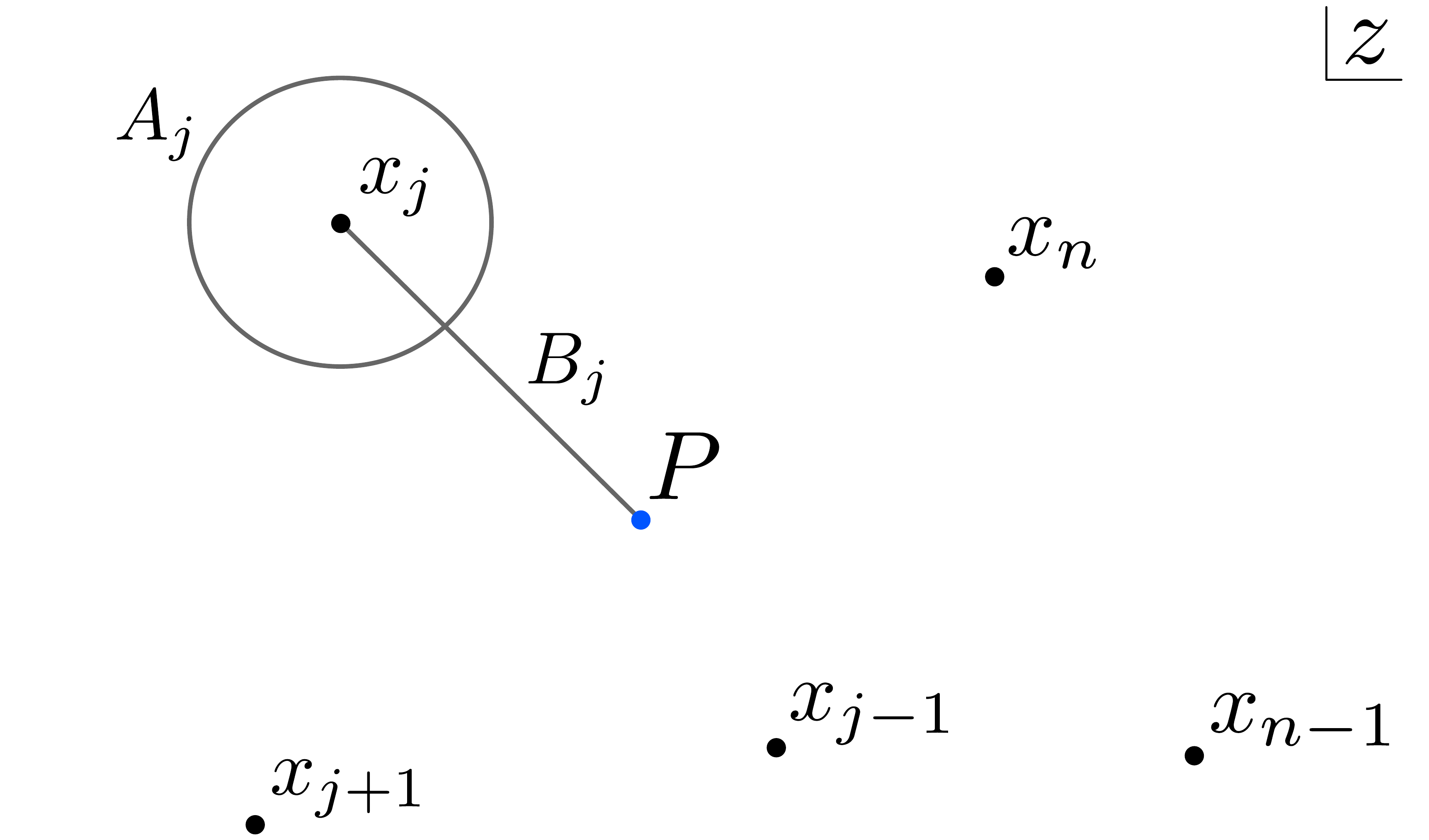}
\caption{\label{PuncturedPlanefig} The $A$- and $B$-cycles of the $n$ punctured $z$-plane.\ The $A$-cycles go around the punctures, whereas the $B$-cycles go from an intersection to a reference point $P$, indicated in blue.}
\end{figure}

On-shell, away from the insertions of local operators, the light-cone scalar fields $X^\pm$ are harmonic functions of $(z,\bar{z})$.
This allows us to simplify this expression further, by making use of the Riemann bilinear identity to write the surface integral over the punctured plane as a sum of product integrals over its cycles
\bea\label{Riemann}
\int\! d^2 z \, \bar{\partial} X^+ \partial X^-\is \sum_{j=1}^n \left( \int_{A_j}\!\! \bar{\partial}X^+ \!  \int_{B_j} \!\! \partial X^- -   \int_{B_j}\!\! \bar{\partial}X^+ \!\int_{A_j} \!\!\partial X^- \right),
\eea
with $A$ and $B$ a canonical basis of cycles, with required unit intersection $\#(A_i,B_j) = \delta_{ij}$ and $\#(A_i,A_j) = \#(B_i,B_j) = 0$. For the plane with $n$ punctures (i.e. holes or cross caps), the  $A$ cycles are loops surrounding the punctures and the corresponding $B$-cycle is a line running from the puncture to a reference point $P$, see figure \ref{PuncturedPlanefig}.\footnote{Finite size holes or cross caps do not have a unique location. Due to the special boundary conditions on the $X^\pm$-fields at the holes or cross caps, one can choose the $B$-cycle to end at any point on the circle $(z-z_i)(\bar{z}-\bar{z_i} )= y_i^2$. }

Next we note that the contour integral over the $A$-cycle can be easily evaluated to
\bea\label{acycleid}
\frac 1 {2\pi } \oint_{A_i}  \bar\partial X^+  \; O(x_i) \is - {\partial}_{-}  O(x_i), \qquad \quad \frac 1 {2\pi} \oint_{A_i} \partial X^-  \; O(x_i) \, = \, {\partial_+}  O(x_i).
\eea
This equation is compatible with the Virasoro relation, which expresses the light-cone coordinate fields in terms of the energy-momentum tensor of the transverse CFT. Similarly, we can use the Virasoro condition to express the $B$-cycle line integral as
\bea
\label{bcycleid}
\int_{B_i}\! \bar\partial X^+ \is \int^{x_i} \!\! \!\! dx^-\,T_{--} , \qquad \quad \int_{B_i}\!  \partial X^-  \, = \,  \int^{x_i} \!\!\!\! dx^+\, T_{++} .
\eea
Here the upper limit $x_i$ is meant as a symbolic notation, indicating that the integral is ending on the finite size `puncture' associated with the local operator $O(x_i)$. Within our formalism, this is the most concrete definition of the line integral of the energy-momentum tensor up to the location $x_i$ of the operator. Note in particular that the $B$-cycle integral on the left-hand side does not need to be regularized: our formalism provides an automatic UV regulator, and does not require the introduction of any other UV scales except for the $T\bar{T}$ coupling $\mu$ itself.

As noted earlier, both equations in \eqref{bcycleid} strictly only hold if the other component of the energy-momentum tensor vanishes, 
since otherwise the lightcone coordinate $x^\pm$ used inside the line-integral no longer behaves classically. 
From the point of view of the worldsheet, equations \eqref{bcycleid} are valid and combining it with \eqref{acycleid}, we can write
\bea
-\frac 1 {2\pi}  \left( \int_{A_i}\!\!\! \bar{\partial}X^+ \! \int_{B_i}\!\!\! \partial X^-\!\! -\!   \int_{B_i} \!\!\! \bar{\partial}X^+ \! \int_{A_i}\!\!\! \partial X^- \right) O(x_i)\, \is  \int^{x_i} \!\! \bigl( dx^+\,T_{--} \partial_- +
  dx^-\,T_{--} \partial_+ \bigr) \, O(x_i) \qquad\nonumber\\[-2mm]\label{cardyflow}
  \\[-2mm]\nonumber
  \is   \epsilon^{ab} \epsilon^{cd} \int^{x_i} \!\!\!\! dx_a\, T_{bc}(x) \, \partial_d O(x_i) \, .
\eea
In the second equality we used the fact that the energy momentum tensor of the CFT is traceless. Going beyond the linearized approximation would involve iterating the flow equation and include the backreaction of the energy momentum on the coordinate system. This integration procedure amounts to expressing everything in terms of target space coordinates $X^{\pm}$. In \cite{Cardy:2019qao} it was argued that the second form of the equation \eqref{cardyflow} then remains true in the full non-linear theory, leading to the general flow equation for correlation functions in the deformed theory
\bea
\frac{\partial} {\partial\mu} \bigl\langle{\, \prod_i O(X_i)}\, \bigr\rangle \, \is \, \sum_{i=1}^n\Bigl\langle \Bigl( \epsilon^{ab} \epsilon^{cd}\! \int^{X_i} \!\!\!\! dX_a\, T_{bc}(X) \, \partial_d O(X_i) \Bigr)\; \prod_{j\neq i} O(X_j) \Bigr\rangle.
\eea

\def\eepsilon{\mbox{\small ${\cal E}$}}

\def\li{|}
\def\ra{\rangle}

\medskip

\section{Three-point function}\label{sec:threepnt} 
\vspace{-1mm}

In this subsection we will describe the computation of the matrix element of a local operator between a given in- and out-state. This calculation can be viewed as the first determination of the $T \bar{T}$ deformation of the OPE coefficient. 
As before, we place the $T\bar{T}$ theory on a cylinder with circumference $2\pi R$, parametrized by coordinates $(x_0,x_1)$ with $0 \! \leq\! x_1 \! <\! 2\pi R$. 
We will choose natural units $|\mu| = 1$. Turning off $\mu$ amounts to sending the radius $R \to \infty$.

Let $| \eepsilon, J \rangle$ denote a state with energy $\eepsilon$ and momentum $J$. We set out to calculate the 3-point function of the deformed theory as a function of $R^2$
\bea
	\langle \eepsilon',J' | \widetilde{\cal O}_{h}(t,\phi)
	| \eepsilon, J \rangle  \is 	c_{123} (R^2)\, e^{-i t \eta} e^{i \phi \ell} 
\eea
with $t= {x_0}/R, \phi = {x_1}/R$, $\eta = \eepsilon' \! - \eepsilon$ and $\ell = J' \! - J \in  \mathbb{Z}$. 
In the limit $R^2\to\infty$, this 3-point function approaches the CFT answer
\bea	
\langle \Delta',J' | \mathcal{O}_h(t,\phi) 
	| \Delta ,J \rangle  \is c_{123}\, e^{-i t \omega} e^{i \phi \ell} 
\eea
where $\omega = \Delta' - \Delta$ equals the difference between the conformal dimensions of the initial and final state, and $c_{123}$ denotes the OPE coefficient of the undeformed CFT. Without loss of generality, we will from now on place the operator at the special location $t=0$ and $\phi=0$. The three point function then reduces to the $T\bar{T}$ deformed OPE coefficient
\bea
	c_{123} (R^2) \is \langle \eepsilon',J' | \widetilde{\cal O}_{h}(0)
	| \eepsilon, J \rangle 
\eea
where  $\widetilde{\cal O}_{h}(0)$ is the cross cap operator placed at the origin  $(t,\phi) = (0,0)$. 

It will be convenient to define the in- and out Hilbert space using radial quantization around two (initially) arbitrarily chosen points $z_1$ and $z_2$. According to \eqref{phystate}, we can write the initial and final state as follows 
\bea 
	| \eepsilon,J \rangle \is e^{i k_1 \cdot X(z_1, \bar z_1)} \,{\cal O}(z_1,\bar z_1) c(z_1) \bar{c}(\bar{z}_1)  \spc |0\rangle\nonumber\\[-2.5mm]\\[-2.5mm]\nonumber	
	\langle \eepsilon',J' | \is \langle 0 | c(z_2) \bar{c}(\bar{z}_2) e^{i k_1 \cdot X(z_2, \bar z_2)} \,{\cal O}(z_2,\bar z_2) 
\eea
where the four chiral space-time momenta $k_i$ are given by 
\bea
\label{kmomenta}
	k_1^\mu \is p_{1L}^\mu \spc = \Bigl(\spc \eepsilon' -  R 
	,  \frac{J'}{R} - R\spc
	\Bigr) , \qquad -k_4^\mu \spc =\spc p_{1R}^\mu \spc= \spc \Bigl(\spc \eepsilon' -  R
	,  \frac{J'}{R} + R\spc
	\Bigr) 	\\[3.5mm]
	-k_2^\mu \is p_{2L}^\mu \spc =\spc \Bigl(\spc \eepsilon -  R
	,  \frac{J}{R} - R
	\spc \Bigr) , \qquad\ \ \ k_3^\mu \spc =\spc p_{2R}^\mu = \Bigl(\spc \eepsilon -  R
	 ,  \frac{J}{R} + R\spc 
	 \Bigr) . 
\eea
Here, for later application, we have chosen the {\it holographic sign} $\mu =-1$ of the $T \bar{T}$ coupling. This holographic sign corresponds to a negative string tension of the worldsheet theory.  

\begin{figure}[t]
	\centering 
	\includegraphics[width=10cm]{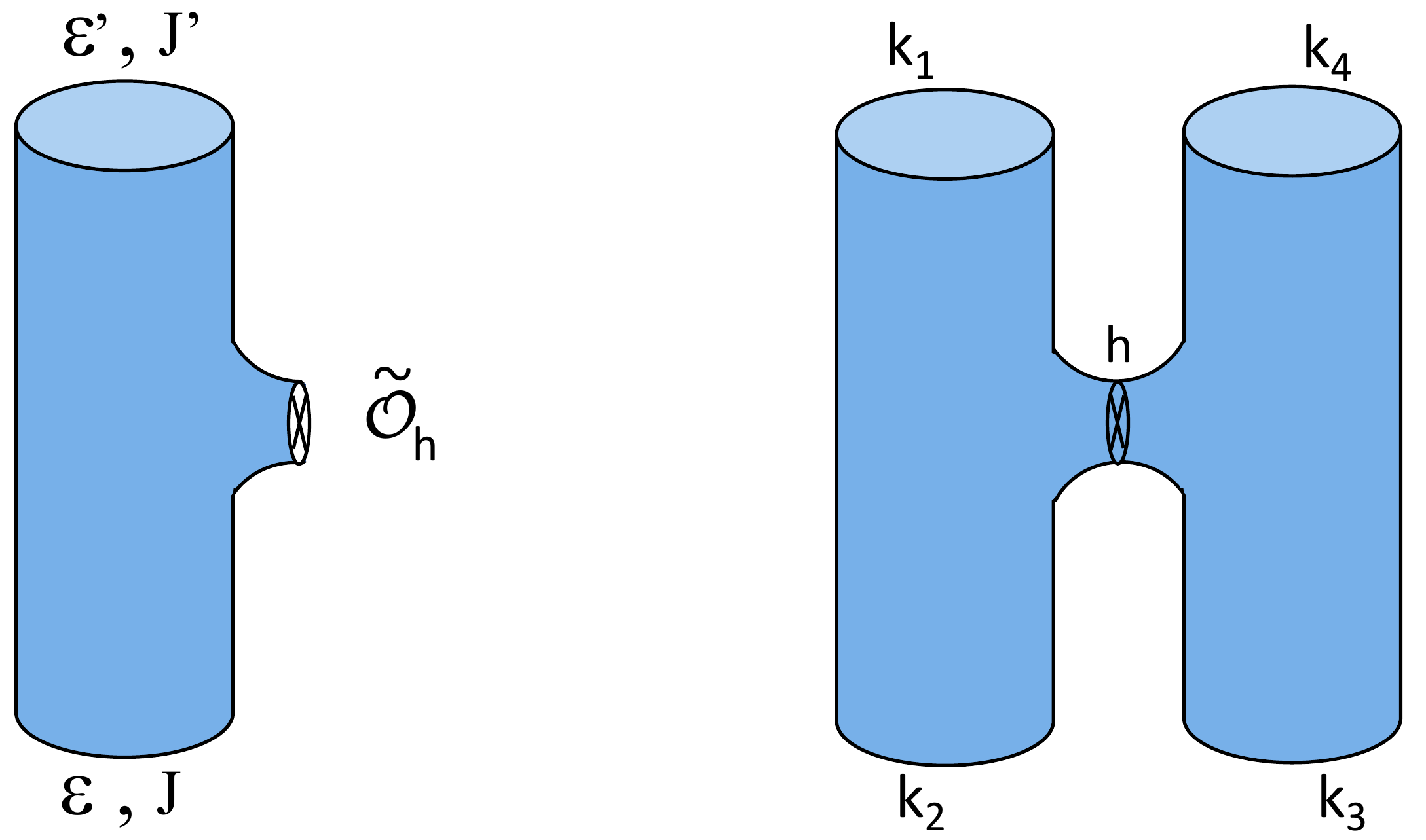} 
	\caption{The local operator $\widetilde{\cal O}_h$ in the deformed theory creates a finite size cross cap. The matrix element between energy-momentum eigenstates can be computed by mapping to the Schottky double. The energy and momentum are encoded via four target space momenta given in equation \eqref{kmomenta}.}
\end{figure}

The computation follows a familiar set of steps. 
For concreteness, we describe here the computation for the cross cap boundary state.
As indicated in Figure 1, the cylinder with a cross cap insertion can be unfolded into a doubled geometry, called the Schottky double, consisting of two cylinders connected by a tube. The original cross cap geometry is obtained by modding out an orientation reversing $\mathbb{Z}_2$ identification.  The doubled geometry has one real complex structure modulus, which can be identified with the real cross ratio $\rho$ of the four punctured sphere with a $\mathbb{Z}_2$ involution.

The effect of the cross cap operator $\widetilde{\cal O}_{h}$ at worldsheet location $(z,\bar z)$ on an operator ${\cal O}(z_1,\bar z_1)$ is to place a virtual, chiral copy at the mirror side of the Schottky double,  ${\cal O}(z_1,\bar z_1) \rightarrow {\cal O}_{\mbox{\tiny ${\nspc \Delta_L}$}}(z_1)$ ${\cal O}_{\mbox{\tiny ${\nspc \Delta_R}$}}(z_1')$, with $z_1' = z - \frac{y^2}{\bar z_1 - \bar z}$ as prescribed by the cross cap identification.  
At this point we can make use of the Mobius invariance to fix three of the four positions to $z_1 \rightarrow \infty$\footnote{The factor $\mathcal{M}$ in \eqref{correction} is crucial here to ensure that there is no divergence as we take $z_1 \to \infty$.}, $z_2 \rightarrow 0$ and $z_2' \rightarrow 1$ and $z'_1 = z$. 
We are left with one integral over the unfixed position $z$  
\bea	
\langle \eepsilon',J' | \widetilde{\cal O}_{h}
| \eepsilon, J \rangle   \is   \\[2mm] & & \hspace{-2cm}
\int_1^{0}\!\!\! dz \; \bigl\langle e^{ik_1X(\infty)}  e^{ik_2X(0)}  e^{ik_3X(1)}  e^{ik_4X(z)}\bigr\rangle\,
\bigl\langle {\cal O}_{\mbox{\tiny ${\nspc \Delta'_L}$}}\!(\infty) \spc {\cal O}_{\mbox{\tiny ${\nspc \Delta_L}$}}\! (0) \spc {\rm P}_{h,p} \spc {\cal O}_{\mbox{\tiny ${\nspc \Delta_R}$}}\!\spc(1) \spc {\cal O}_{\mbox{\tiny ${\nspc \Delta'_R}$}}\!(z)
\bigr\rangle . \nonumber 
\eea
Here ${\rm P}_{h,p}$ is the projection operator onto the chiral sector spanned by all descendents (defined as the states obtained by acting with Virasoro generators and free field $X$ oscillators)  of the primary state $\li h\ra \li p \ra$ with conformal dimension $h$ and momentum $p$.  
Converting to an integral over the real modulus $\rho \equiv \frac{z_{12}z_{34}}{z_{14}z_{32}} 
= \frac{y^2}{z \bar z+y^2} 
$  of the Schottky double geometry, we finally obtain 
\bea 
\langle \eepsilon',J' | \widetilde{\cal O}_{h}
| \eepsilon, J \rangle  \; = \;  
\int_0^1\!\! d\rho \, f(\rho) \,\bigl\langle {\cal O}_{\mbox{\tiny ${\nspc \Delta'_L}$}}\!(\infty) \spc {\cal O}_{\mbox{\tiny ${\nspc \Delta_L}$}}\! (0) \spc {\rm P}_{h,p} \spc {\cal O}_{\mbox{\tiny ${\nspc \Delta_R}$}}\!\spc(1) \spc {\cal O}_{\mbox{\tiny ${\nspc \Delta'_R}$}}\!(1-\rho)
\bigr\rangle  
\label{veneziano}
\eea
with $f(\rho)$ the vertex operator correlation function computed using the techniques outlined in section \ref{Xpmcorrelators}. Using the free field answer combined with the general expression \eqref{correction} for the $\kappa$ dependent conformal anomaly contribution, we find that the $T \bar{T}$ deformed correlation function is obtained by integrating a CFT conformal block over a dressing kernel
\bea
f(\rho) \is  \frac{1}{(k_{1}^+ k_{2}^+ k_3^+k _4^+)^{\kappa/2}}
 \left(\frac{\rho^{2}(1-\rho)^{2}}{(k_2^+ \rho - k_3^+ (1 \! -\! \rho) + k_4^+)^2 + 4 k_3^+ k_4^+ (1\! -\! \rho)}\right)^{\!\spc \kappa/2} \! \rho^{-\frac{1}{2} k_3 \cdot k_4} (1-\rho)^{-\frac{1}{2} k_2 \cdot k_4} \, .\nonumber \\[-3mm]
\eea

\subsection{Check of holographic dictionary}

In the proposal of \cite{MMV}, turning on the $T \bar{T}$ coupling with $\mu<0$ amounts to moving the CFT into the bulk (see also \cite{Hartman:2018tkw, Kraus:2018xrn, Taylor:2018xcy, Guica:2019nzm, Gross:2019ach}). In this case, it indeed is most natural to choose the cross-cap definition of the local operator. 
{According to \cite{ooguri,HV,taka} we can interpret a global cross-cap state with some finite size as the HKLL representation of a local bulk field placed at a finite radial location in the AdS bulk. In \cite{HV,Lewkowycz:2016ukf} it was proposed that the gravitationally dressed operators are naturally given by the full cross-cap Ishibashi state. This leads us to suspect that the given prescription for the deformed correlation function can be used as a non-trivial test of the holographic interpretation of the $T \bar{T}$  deformation in \cite{MMV}. }

Let $y$ be the radial $AdS_3$ coordinate in the Poincar\'e patch 
\bea
ds^2 \is \frac{ dy^2 + dzd\bar z}{y^2}.
\eea
As pointed out in \cite{HV,ooguri,taka}, global conformal transformations 
that commute with the cross-cap identification 
\bea 
x-z  \is - \,\frac {y^2}{\bar{x} -\bar{z}}
\eea
  map to space-time isometries that leave the corresponding bulk point $(y, z,\bar{z})$  in  invariant. Note, however, that in our context the size of the cross-cap is not fixed, but represents a modulus that we should integrate over.

\def\ZZ{\mbox{\small $Z$}}

Intuitively, we expect that for large $R^2$, or equivalently, for small $T\bar T$ coupling, the dressing factor is a sharply peaked function with a maximum $\rho_c(R^2)$ that approaches $ \rho_c \to 0$ as $R^2 \to \infty$. This behavior ensures the correspondence with the undeformed theory. 

\begin{figure}[t]
	\centering \includegraphics[scale=1.05]{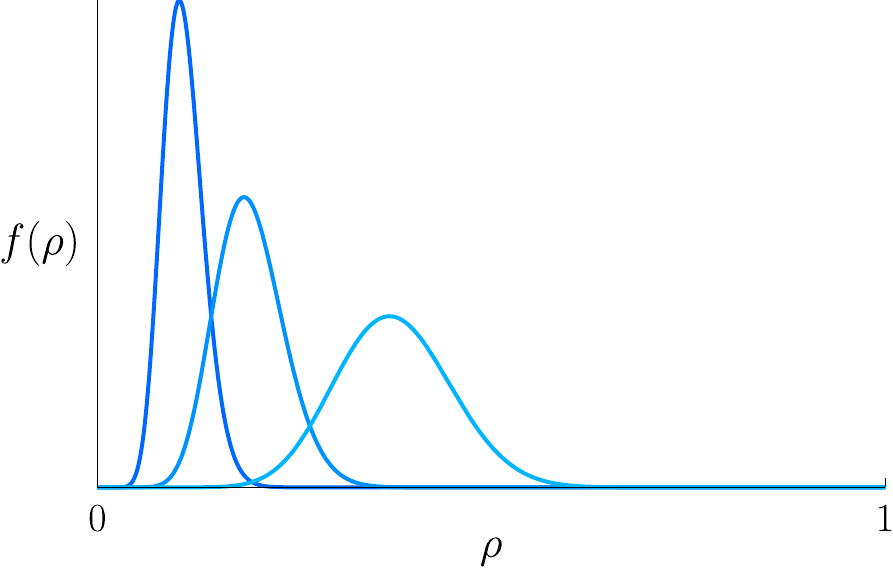} 
	\caption{Plot of normalized smearing function $f(\rho)$ in the limit of large $R$ and $\Delta$ for decreasing values of $R$  (decreasing shade of blue). 
	}
	\label{figsmearing}
\end{figure}

Here we will perform a quantitative check of this physical expectation in the regime where the initial and final states have energy ${\cal E}\gg\frac{c}{ 12 R}$. For simplicity, we will also set the angular momentum $J=J'=0$. In this regime and for large $R$, the exponents of $\rho$ and $(1-\rho)$ in the dressing prefactor $f(\rho)$ in \eqref{veneziano} take the approximate values
\bea
	-\frac{1}{2} k_3\nspc\cdot\nspc k_4 \is \, \frac{1}{2}R(\eepsilon +\spc \eepsilon' )- \spc \frac{1}{2} {\eepsilon\eepsilon'}\,  \quad \quad \ \ \raisebox{-1.4mm}{${\longrightarrow}\atop{\raisebox{.5mm}{\tiny $R$  large}}$}\; \, \; M    \label{powerofrho}  \\[2mm]
	-\frac{1}{2} k_2\nspc\cdot\nspc k_4 + \kappa \is R^2\! - \frac{1}{2}R( \eepsilon + \eepsilon')\! + \frac{1}{2} {\eepsilon\eepsilon'} + \kappa  \; \ \raisebox{-1.4mm}{${\longrightarrow}\atop{\raisebox{.5mm}{\tiny $R$  large}}$}\; \, \; R^2 - M 
\eea
with $M \simeq R \eepsilon \simeq R \eepsilon'$. Hence we see that for large $R^2$,  $f(\rho)$ is indeed a sharply peaked function with a maximum at
\bea
\label{rhoc}
\rho_c 
\, \simeq\, \frac{M}{R^2}.
\eea 
This is illustrated in Figure \ref{figsmearing}, where the leading $R$ behavior of the smearing function $f(\rho)$ is plotted.

Let us compare this with the proposed holographic dictionary \cite{MMV}.  The bulk space time dual to the state with mass $M$ is described by the BTZ black hole metric
\bea
ds^2 \is -\Bigl(1 - \frac{r_+^2}{r^2} \Bigr)dt^2 + \frac{dr^2 }{1- r^2_+/r^2}  + r^2 d\phi^2\eea
with $r_+^2 =8G_NM = \frac{12M}{c}$. On the CFT side, we note that at large central charge  $c$, a CFT amplitude of heavy operators selects a special `uniformizing' coordinate system $(\ZZ,\bar{\ZZ})$, such that the expectation value of the stress tensor vanishes: $\langle T(\ZZ)\ra = \langle \bar{T}(\bar{\ZZ})\ra =  0$.  Such a coordinate system always exists, thanks to the anomalous transformation rule of $T(z)$. The behavior $\langle T(z)\ra={ \Delta}/{z^2}$ near a heavy operator with dimension $\Delta$ is uniformized by  
\bea
\label{uniformz}
\label{rplus}
\ZZ(z)= z^{ir_{\! +}} \qquad {\rm with} \qquad r_{\!+}^2 =  {\frac{24 \Delta}{c}-1} = \frac{12 M}{c}.
\eea 
The $(\ZZ,\bar{\ZZ})$ coordinates are multivalued: under $z \to e^{2\pi i} z$, they undergo a monodromy specified by the same hyperbolic $SL(2,\mathbb{R})$ elements that characterize the corresponding BTZ geometry.  

The above observations combined indicate that the cross ratio $\rho$ of the Schottky double geometry must be identified with the radial holographic coordinate $r$ via
\bea
 \rho \spc\is \spc \frac{r_+^2}{r^2} .
\eea
Combining this with equations \eqref{rhoc} and \eqref{rplus}, we find that the location $\rho = \rho_c$ of the peak in the smearing  $f(\rho)$ corresponds to the radial bulk location
\bea
  r^2_c \spc = \spc \frac{r_+^2 }{\rho_c} \spc \simeq \spc \frac{12 R^2}{c} .
\eea
This precisely matches with the proposed holographic dictionary put forward in \cite{MMV}. 

\medskip

It is perhaps tempting to make use of a saddle point approximation of the integral in \eqref{veneziano} to conclude that the deformed correlation function is approximated by the undeformed correlation function evaluated at the cut-off radial location $\rho =\rho_c$. This does not quite work out, however, because the conformal block itself also has a steep dependence on $\rho$ near $\rho =0$. The small $\rho$ behavior of the combined integrand is governed by the lightest state that propagates through the intermediate channel, i.e. the small bridge that connects the two sides of the Schottky double. This state has total scale dimension $L_0 = h + (k_1+k_2)^2 = h - (\eepsilon - \eepsilon')^2$.  So at small $\rho$, the integral behaves as $\int \! d\rho\, \rho^{L_0 -1} = \int \! d\rho \, \rho^{h - (\mbox{\tiny ${\cal E} - {\cal E}'$})^2 -1}$.  Note that the $\rho$ integral converges as long as $h > (\eepsilon - \eepsilon')^2 $. In the other regime $h < (\eepsilon - \eepsilon')^2$, the amplitude needs to be defined via analytic continuation.

\section{Discussion}

In this paper we have given a proposal for a non-perturbative definition of a $T\bar{T}$ deformed field theory. Our proposal entails the coupling of the original CFT to a two dimensional non-critical string theory worldsheet in such a way that the total central charge of the worldsheet theory vanishes. This allows for a rigid definition of the Hilbert space in terms of the BRST cohomology. In section \ref{sec:spectrum}, we showed that the spectrum and thermal partition function obtained using our formalism indeed matches with the one expected from known results in the unit winding sector. Most importantly, our proposal gives a recipe for how to construct local operators of which we have studied two types either using cross-caps or D-branes. We used these local operators to construct correlation functions and find a flow equation for them. In section \ref{sec:threepnt} we computed a three-point function in the deformed theory and compared to the holographic interpretation of the $T\bar{T}$ deformation. 
There are of course many unanswered questions and interesting future directions. 

\subsection*{\it Open strings?}
In the cross-cap computation in section \ref{sec:threepnt} the worldsheet is the twice punctured real projective plane, which does not have any boundaries. Hence from the non-critical string perspective, only closed strings scatter in this process.  When we consider normal Ishibashi states, the geometric identification of the coordinates does have fixed points, resulting in a true boundary of the string worldsheet. This signals the presence of open strings that can end on that boundary. Indeed, as mentioned in the text, such boundary states would be D-instantons. A simple observable involving this particular boundary condition is the cylinder amplitude, the D-instanton two-point function. In the closed string picture, this would be a closed string emitted and absorbed by a D-instanton. For a general Ishibashi boundary state, this amplitude takes the form 
\bea
A(x_1,x_2) \is \frac{1}{\mathcal{N}_{D(-1)}^2} \int_0^{\infty} \!\! \frac{dl}{4\pi^2 l} \; e^{-\frac{1}{4\pi l}(x_1 - x_2)^2}\chi_h(2i l)
\eea
with $\chi_h(2i l)=\braket{h|e^{-2\pi l (L_0^{\rm CFT} + \bar{L}_0^{\rm CFT}-(c+\bar{c})/24)}|h}$. These amplitudes do a priori not correspond to anything known in the undeformed theory, but are valid observables in the deformed theory. It would be interesting to study these new types of observables in more detail.

\subsection*{\it Towards entanglement entropy in string theory}

Given that we have shown an equivalence between non-critical string theory and a $T\bar{T}$ deformed conformal field theory, it would be interesting to study entanglement entropy in this context. Entanglement entropy is usually computed using the replica trick, which, from our perspective, would mean a replica trick on the worldsheet. As a result of the coordinate transformation from the worldsheet to the target space, this will also result in a replicated target space geometry. Ideally, this computation should yield a finite entanglement entropy, similar to what we saw happening to the three-point function in section \ref{sec:threepnt}. A particularly interesting question is whether, due to the restriction to the winding one sector and hence the absence of $T$-duality, how and if the conical singularity in target space gets resolved or not. 

\subsection*{\it A new non-critical string theory?}

The non-critical string theory we proposed as a non-perturbative description of the $T\bar{T}$ deformed theory is also interesting in its own right. Our theory can be seen as a 2D version of the Polchinski-Strominger non-critical string \cite{Polchinski:1991ax}, which like ours is a Lorentz-invariant non-linear theory. A special property of the 2D theory is that the non-linear terms can be simply incorporated via an overall measure factor \eqref{correction}. This fact was known already in the old string theory literature \cite{green1988superstring}. This old formulation of non-critical string theory should be contrasted with the (recently more standard) formulation of non-critical string theory with a linear dilaton background, such as discussed e.g. in \cite{Dodelson:2017emn}. The linear dilaton background breaks Lorentz invariance along the light-cone directions, and thus looks different from ours. The two formulations are related, however, via the field redefinition presented in section 2. This field redefinition can be seen as a reinterpretation of how the target space geometry is encoded inside the linear dilaton theory. It would be worthwhile to study this mapping between the two different formulations of non-critical string theory in more detail.

\subsection*{\it JT gravity and other Planck branes}

It is natural to ask if our construction can be generalized to define a modified $T \bar{T}$ deformed theory by coupling the CFT to the JT gravity action with non-zero cosmological constant. Unlike our $X^\pm$ theory and the flat dilaton gravity model, JT gravity is a truly non-linear theory, albeit one without local degrees of freedom. So it should still be possible to choose an analogue of the time-like gauge and use the uniformizing $AdS_2$ coordinate system as  worldsheet coordinates. The effective theory in this gauge will look like a CFT with a non-linear generalization of a $T\bar{T}$ deformation. Moreover, it seems natural to assume that this deformed CFT will have an intrinsic UV cut-off. 

By applying the holographic mapping, this coupled CFT+JT gravity system can be given an effective description in terms of a compactified $AdS_3$ space-time, with the asymptotic region replaced by an effective Planck brane. This exact type of situation was recently used in \cite{Almheiri:2019hni} to study the dynamics of quantum extremal surfaces in 2D CFT coupled to JT gravity. From the 3D perspective, the appearance of JT gravity is an example of the RS scenario \cite{Randall:1999ee}, i.e. the localization mechanism by which a radial cutoff leads to dynamical gravity along the Planck brane. Conversely, this dictionary provides additional support for the proposed holographic interpretation of the $T\bar{T}$ deformations and its non-linear JT gravity generalization. 

Another context where a coupled CFT+JT theory was obtained from integrating out UV degrees of freedom is provided in the work in \cite{Callebaut:2018nlq,Callebaut:2018xfu}. Namely, it was found that entanglement degrees of freedom of the CFT could be treated as a collective mode of the CFT, described by an effective CFT+JT action. It would be interesting to better understand the connection to $T \bar{T}$ in that setting as well.

\section*{Acknowledgements}

We want to thank Jan de Boer, David Gross, Emil Martinec, Edward Mazenc, Thomas Mertens,  Shiraz Minwalla, Vladimir Rosenhaus, Eva Silverstein, Ronak Soni and Edward Witten for useful discussions and comments. N.C. was supported in part by the Research Foundation-Flanders (FWO Vlaanderen) and in part by the Israel Science Foundation under grant 447/17. JK is supported by the Simons Foundation. The research of H.V. is supported by NSF grant PHY-1620059.

\appendix
\section{OPE's with $T$}\label{appA}

In this appendix we outline the computation of OPE's involving the $\kappa$-dependent stress tensor \eqref{tnew}. 

\subsection{OPE between stress tensors} 

Using the standard Wick contraction rule \eqref{wick}
one finds that the contribution to $T(z)T(w)$ proportional to $\kappa$ takes the form 
\bea 
& & \kappa   \left \{ \p_z^2 \left( \wick{\c{\log\bigl( \p X^+(z)\bigr)} \c{\p X^-}(w)} \right) \p X^+(w) + \p X^+(z) \, \p_w^2 \left( \wick{\c{\p X^-}(z) \c{\log \p X^+(w})}\right) \right \}  \nonumber \\[2.5mm]
& & \ \ \ = - \kappa \left \{ \p_z^2 \left( \frac{1}{\p X^+(z)}  \frac{1}{(z-w)^2} \right) \p X^+(w) + \p X^+(z) \, \p_w^2 \left(  \frac{1}{\p X^+(w)}  \frac{1}{(z-w)^2} \right) \right \}. 
\eea 
In the second line we used the chain rule to evaluate the contraction between the logarithm of $\p X^\pm$ and $\p X^\mp$. It remains to perform a Laurent series expansion around $z=w$ of this expression, which gives 
\bea  
- \frac{12 \kappa\ }{(z-w)^4} - \frac{2 \kappa \, \p^2 \log \p X^+(w)}{(z-w)^2} - \frac{\kappa \, \p^3 \log \p X^+(w)}{z-w} + \cdots
\eea 
as required to find the result in \eqref{TTOPE}. 

\subsection{OPE between stress tensor and plane wave operator}

Here we give some details of the computation of the OPE between the modified stress tensor \eqref{tnew} and plane wave operators. In fact, we can make a more general statement. Let us consider the OPE between $\mathcal{F}(\partial X(z))$ and $e^{ipX(w)}$. Without loss of generality we omitted the index structure on $X$ and $p$ here. The contraction of $X$ with itself is 
\bea
\wick{\c{X(z)}\c{X(w)}} \is -\log(z-w) \, .
\eea
Let us write $\mathcal{F}(\partial X(z))$ as a power series in $\partial X$. The OPE we are then interested in is
\begin{align}
:(\partial X(z))^n:: e^{ipX(w)}: &\sim \sum_{k=1}^n \begin{pmatrix} n \\ k \end{pmatrix} \frac{(ip)^k}{(z-w)^k} (\partial X(z))^{n-k} e^{ip X(w)}\nonumber\\[-2.5mm]\\[-2.5mm]\nonumber
&= \left[\left(\frac{ip}{z-w} + \partial X(z)\right)^n - (\partial X(z))^n \right]e^{ip X(w)} \, .
\end{align}
Hence,
\bea
:\mathcal{F}(\partial X(z)):: e^{ipX(w)}: & \sim & \left[\mathcal{F}\left(\frac{ip}{z-w} + \partial X(z)\right) - \mathcal{F}(\partial X(z)) \right]e^{ip X(w)}.
\eea
Now we should expand the $\partial X$ around $z = w$ to get the final form of the OPE. It is clear that many derivatives of $X$ at $z = w$ can appear in that way and so to make things concrete, let us apply it to our situation in which the non-trivial OPE is with $\mathcal{F} = -\k \partial^2 \log(\partial X(z))$. Expanding around $z = w$ then gives
\bea
-\k : \partial^2 \log(\partial X(z))::e^{ip X(w)}: & \sim & -\frac{\k}{(z-w)^2} + (\rm regular ),
\eea
which shows indeed that the $\k$ correction to the stress tensor gives rise to a shift in the conformal dimension by $-\k$. It is interesting to note that stress tensors with a correction proportional to the Schwarzian derivative of $X$ as appeared in the literature before \cite{Baba:2009ns} do \emph{not} give rise to this shift. In fact, the OPE of the plane waves with such a correction is regular.
\small
\bibliographystyle{JHEP}
\bibliography{referencesTTbar}

\end{document}